\documentclass[aps,pre,twocolumn,showpacs]{revtex4}

\usepackage{epsfig,latexsym}
\usepackage{amsmath,amscd,amssymb}
\usepackage{dcolumn}

\newcommand{\be}{\begin{equation}}
\newcommand{\ee}{\end{equation}}
\newcommand{\bea}{\begin{eqnarray}}
\newcommand{\eea}{\end{eqnarray}}

\def\eq#1{Eq.~(\ref{#1})}
\def\eqs#1#2{Eqs.~(\ref{#1})-(\ref{#2})}
\newcommand{\bnabla}{{\bf \nabla}}
\newcommand{\br}{\mathbf{r}}
\newcommand{\bq}{\mathbf{q}}
\newcommand{\bR}{\mathbf{R}}

\newcommand{\bu}{\mathbf{u}}
\newcommand{\bv}{\mathbf{v}}

\begin{document}

\title[Theory of two supported stacked lipid bilayers]{Statistical mechanics and dynamics of two supported stacked lipid bilayers}
\author{Manoel Manghi}
\email{manghi@irsamc.ups-tlse.fr}
\author{Nicolas Destainville}
\affiliation{Universit\'e de Toulouse; UPS; Laboratoire de Physique Th\'eorique (IRSAMC); F-31062 Toulouse, France, and\\CNRS; LPT (IRSAMC); F-31062 Toulouse, France}

\begin{abstract}
The statistical physics and dynamics of double supported bilayers are studied theoretically. 
The main goal in designing double supported lipid bilayers is to obtain model systems of biomembranes: the upper bilayer is meant to be almost freely floating, the substrate being screened by the lower bilayer. 
The fluctuation-induced repulsion between membranes and between the lower membrane and the wall are explicitly taken into account using a Gaussian variational approach.
It is shown that the variational parameters, the ``effective'' adsorption strength and the average distance to the substrate, depend strongly on temperature and membrane elastic moduli, the bending rigidity and the microscopic surface tension, which is a signature of the crucial role played by membrane fluctuations.
The range of stability of these supported membranes is studied, showing a complex dependence on bare adsorption strengths. In particular, the experimental conditions to have an upper membrane slightly perturbed by the lower one and still bound to the surface are found. Included in the theoretical calculation of the damping rates associated with membrane normal modes are hydrodynamic friction by the wall and hydrodynamic interactions between both membranes.
\end{abstract}

\maketitle

\section{Introduction}

\begin{figure}[t]
\begin{center}
\includegraphics[width=8cm]{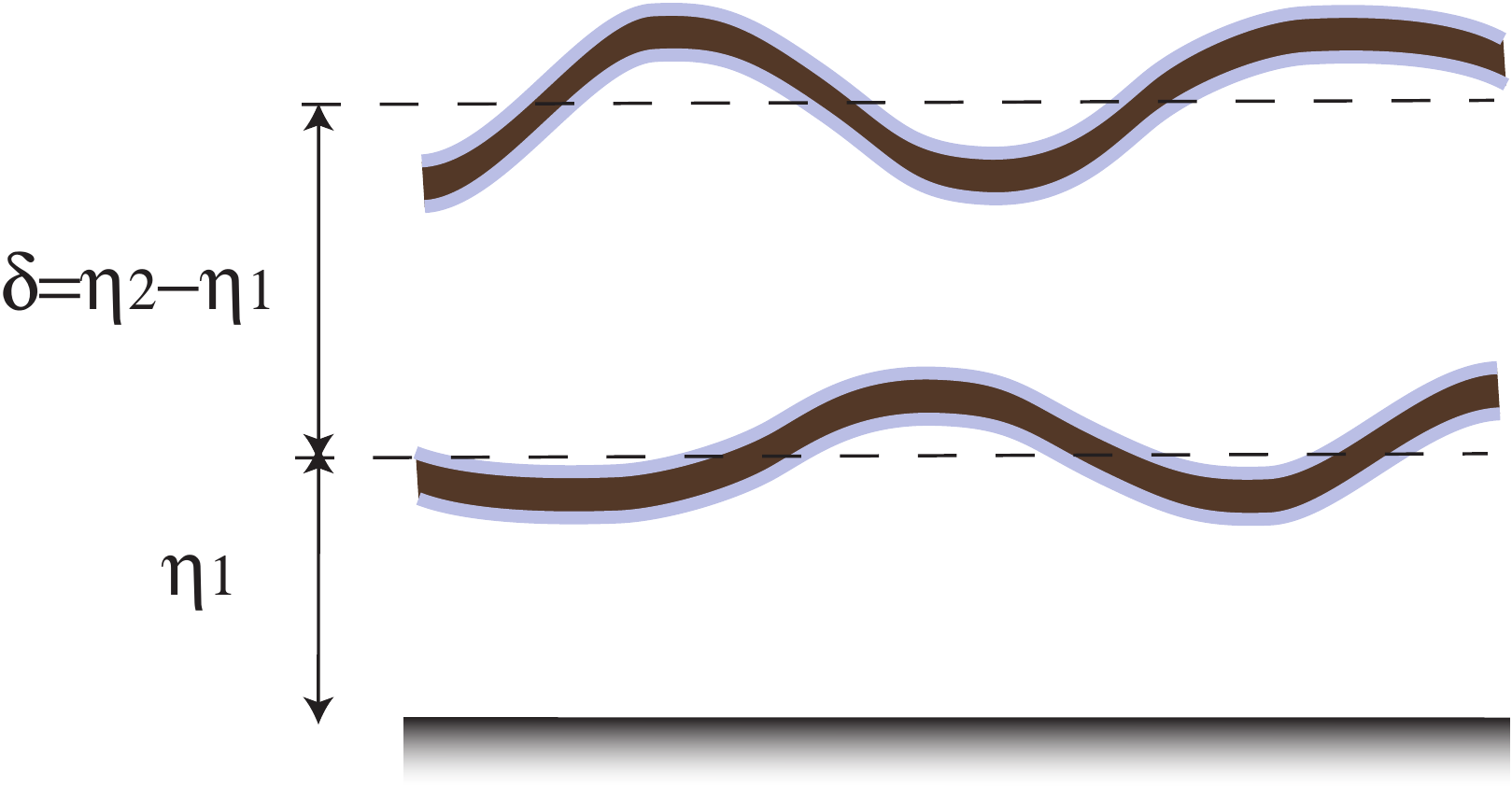}
\caption{Sketch of a double supported lipid bilayer.}
\label{sketch}
\end{center}
\end{figure}

Double supported membranes  are composed of two lipid bilayers superimposed on a solid substrate (\ref{sketch}). The lower membrane (closer to the substrate) is weakly absorbed, the upper one being bound \textit{via} inter-membrane interactions. This system has a growing interest for physicists and biologists since it allows one to avoid, in part, the direct influence of the substrate on the upper membrane (pinning by defects, repulsion or direct attraction by the wall) which becomes almost freely floating. Some recent experimental techniques allow the formation of such systems whether by Langmuir-Blodgett deposition~\cite{daillant,lecuyer,graner} or by deposing after rupture a giant vesicle on a single bilayer~\cite{groves1,groves2}. These double supported lipid bilayers can then play the role of model membranes, the lipid and protein composition of which can be varied. Indeed, the extreme complexity of cell membranes motivates the development of artificial ones~\cite{sackmann,mouritsen} that are more easily studied from a physical point of view. In the present case, the planar geometry facilitates the use of modern spectroscopy techniques to characterize these model membranes~\cite{groves1,daillant,groves2,groves3}. Furthermore, two stacked membranes can model cell-cell junctions~\cite{kaizuka,lin} where the role of lipids and proteins can be investigated~\cite{groves2,partha}.

Since liquid membranes such as lipid bilayers are highly fluctuating at room temperature (their bending modulus $\kappa$ is on the order of several $k_BT$)~\cite{membranes_book}, a fluctuation-induced repulsion appears at finite temperature, leading to a destabilization of the system and eventually, an unbinding from the substrate at a critical temperature $T_u$. 
The unbinding transition of membranes is a long standing issue, first studied by Helfrich~\cite{helfrich2}, which has led to a considerable amount of theoretical investigations on homogeneous stacks of two or more membranes. Lipowsky describes, in a  very nice review~\cite{lipowsky_review}, the various situations where unbinding occurs, among which are the cases of only steric repulsive interactions with applied external pressure, or attractive interactions between membranes, depending on experimental conditions. 

These theoretical studies essentially focus on the order of the transition and the values of the critical exponents, mostly using Monte-Carlo numerical simulations~\cite{zielinska,netz}, or group renormalization techniques~\cite{lipowsky_leibler,lipowskyEPL,grotehans}.  Interestingly, an analytical solution can be obtained when considering the unbinding of fluctuating strings (one dimensional objects) in two dimensions~\cite{burkhardt}. Indeed, this case is exactly mapped onto the delocalization transition of a quantum particle in an external potential and the unbinding temperature $T_u$  can been computed exactly by solving  the Schr\"{o}dinger equation by transfer matrix techniques.

Among these studies one can distinguish between symmetric systems (two identical membranes or a bunch of identical membranes), and asymmetric systems where the membranes are not identical--and eventually the extreme case of one membrane being a solid substrate (formally a membrane with infinite rigidity). Double supported membranes belong to the latter category. The physics of asymmetric stacks studied by Monte-Carlo simulations~\cite{netz,netz2}, or by analytical tools for one-dimensional strings~\cite{hiergeist}, turned out to be very rich, showing for instance, over a certain range of the parameters, a peeling process where successive unbinding temperatures appear, the upper membrane unbinding first. This has to be compared to the case of symmetric stacks where $T_u$ is independent of the number of stacked membranes. However, not much work has been done for real two-dimensional membranes embedded in a tridimensional space. 

Moreover, the major part of these studies consider the effect of bending rigidity in height-height fluctuations, without considering the microscopic surface tension $\sigma$, which might  play an important role in supported bilayers. In such systems, $\sigma$  is related to the chemical potential of the amphiphilic molecules (an energetic parameter associated with the full membrane area and not the projected one)~\cite{david}. The value of this parameter is measured by fitting X-ray reflectivity spectra~\cite{daillant}, where it must be taken into account for wave-vectors $q<\sqrt{\sigma/\kappa}$, and in flickering experiments~\cite{brochard}. Moreover, according to the preparation method, the presence of pinning defects can induce lateral tension~\cite{lin}. 

The surface tension always dominates bending rigidity at large distances, and thus drastically reduces the membrane roughness. It leads to a fluctuation-induced interaction which decays exponentially at large distances, instead of a power law in $z^{-2}$ in the bending rigidity only case~\cite{netzEPL}. By comparing it to the direct interaction, such as van der Waals attraction and screened electrostatic attraction, we thus enter in the weak- or intermediate-fluctuation regimes. It has been shown, using renormalization techniques that, in theses regimes, the order of the transition can change and discontinuous transitions might occur~\cite{lipowskyEPL,grotehans,grotehans2}. 

The techniques presented above which include Monte Carlo simulations, renormalization group calculations, and numerical transfer matrix methods prove successful for calculating quantitative values for critical exponent, but they are not always easy to implement. Moreover, they often do not enhance our intuitive and qualitative understanding of the problem. In this paper, we develop a variational approach of asymmetric unbinding where the inter-membrane and substrate-membrane potentials are modeled by Morse potentials. The variational approach provides an approximation that is analytically simple to implement and can provide a direct link between quantitative calculations and qualitative pictures of the physics of the problem. This analytical work has thus similarities with the ones on single membrane unbinding, which use self-consistent~\cite{helfrich2,evans,manciu,mecke} and variational~\cite{podgornik} calculations. Our goal is to understand the physical mechanisms and the role of the key physical parameters (adsorption potential, bending rigidity, surface tension) in the thermodynamics of supported stacked lipid membranes. What are the parameter ranges for which double supported bilayers are stable?
In particular, non-linearities of potentials are taken into account by the variational parameters, which are effective disjoining pressures and inter-membrane distances, and their dependence with temperature is explicitly studied.

Fluctuating membranes are often studied by light-scattering experiments, which yield information about the bilayer dynamics and the associated damping rates~\cite{seifert}. Moreover, bio-membrane dynamics is crucial for the study of the diffusion of integral membrane proteins~\cite{naji1}, which is influenced both by fluctuations dynamics (projection of the motion onto a reference plane~\cite{reister,gov}) and hydrodynamics~\cite{naji2}. The dynamical fluctuations of a single lipidic membrane in a viscous liquid had been studied in the early seventies by Kramer~\cite{kramer}, who showed that the damping rate is driven by the ratio between viscous damping and bending energy. The presence of an external ``obstacle'', such as a second membrane~\cite{brochard,nelson} or a solid substrate~\cite{seifert_dynamics}, modifies substantially the damping rate as soon as they are close enough to induce hydrodynamic interactions. Non-monotonic behaviour has been observed in the last case. In this work, we generalize these results to double supported lipid bilayers and show how temperature and adsorption potentials modify the wave-vector dependence of damping rates. 

The first Section presents the variational approach in the simplest case of one supported bilayer and a comparison is made with previous works. This variational approach is then applied to two stacked supported bilayers, where phase diagrams and correlation function for height-height membrane fluctuations are computed. Finally, these variational parameters calculated in equilibrium serve to describe the normal modes and damping rates of the system. The flow velocity field is established for the first time in this geometry and hydrodynamic friction and hydrodynamic interactions between the two membranes turn out to be central to understanding this complex dynamics. A discussion of our results is given in conclusion.


\section{One membrane supported on a solid substrate}


We first consider a single membrane lying on a horizontal solid planar substrate at position $z=0$. Membrane local position vector is, in the Monge representation, ${\bf R}=x\hat{\bf x}+y\hat{\bf y}+ h(x,y)\hat{\bf z}$ where $h(x,y)$ is the height of the membrane and  $\eta=\langle h(x,y)\rangle$ is the average distance between the substrate and the membrane. 

The Hamiltonian of a supported membrane is given by the usual Helfrich Hamiltonian of a single fluctuating membrane~\cite{helfrich1,seifert} plus a term taking into account the interaction between the membrane and the substrate:
\be
\mathcal{H}= \int_\mathcal{S} \mathrm{d}^2\br \left\{ \frac12\left[\sigma \left(\bnabla h(\br) \right)^2 + \kappa \left(\bnabla^2 h (\br) \right)^2\right] + V[h(\br)] \right\}
\label{helfrichH}
\ee
where $\br=x\hat{\bf x}+y\hat{\bf y}$ is the projection  of ${\bf R}$ in the plane parallel to the substrate, $\bnabla=\hat{\bf x}\partial_x+\hat{\bf y}\partial_y$ and $\mathcal{S}$ is the projected membrane area. The membrane bending rigidity is $\kappa$ ($\simeq20-50 k_BT$ for a lipid bilayer) and $\sigma$ is the microscopic membrane surface tension.

Several contributions arise in the potential $V(z)$: at short distances it is dominated by excluded volume interactions and  hydrophobic or hydration repulsion (entropic interaction associated to the presence of water molecules inserted between hydrophilic lipid heads)~\cite{evans} . At intermediate distances, it is dominated by screened electrostatic and van der Waals interactions which are essentially attractive.
More precisely, it is usually admitted that, if the Debye screening length is small, the potential between the substrate and the membrane, assumed flat, is essentially the sum of two terms~\cite{lipowsky_review}: a van der Waals attractive part between a flat membrane of thickness $\epsilon$ and an semi-infinite medium~\cite{parsegian_book}
\be
V_{\rm vdW}(z)=-\frac{A_{\rm H}}{12\pi}\left[\frac1{z^2}-\frac1{(z+\epsilon)^2}\right]
\label{vdW}
\ee
where $A_{\rm H}$ is the Hamaker constant between interfaces substrate/water and lipid/water ($A_{\rm H}\simeq1-10\times10^{-21}$~J), and a decreasing exponential repulsive part which takes both steric and hydration contributions into account
\be
V_{\rm hyd}(z)= P_{\rm hyd} e^{-z/\lambda_{\rm hyd}}
\label{hyd}
\ee
where the values of the hydration pressure $P_{\rm hyd}$ ($1-10\times10^{-2}$~J.m$^{-2}$) and $\lambda_{\rm hyd}$ ($0.1-0.3$~nm) are not precisely known.

\begin{figure}[t]
\begin{center}
\includegraphics[width=8cm]{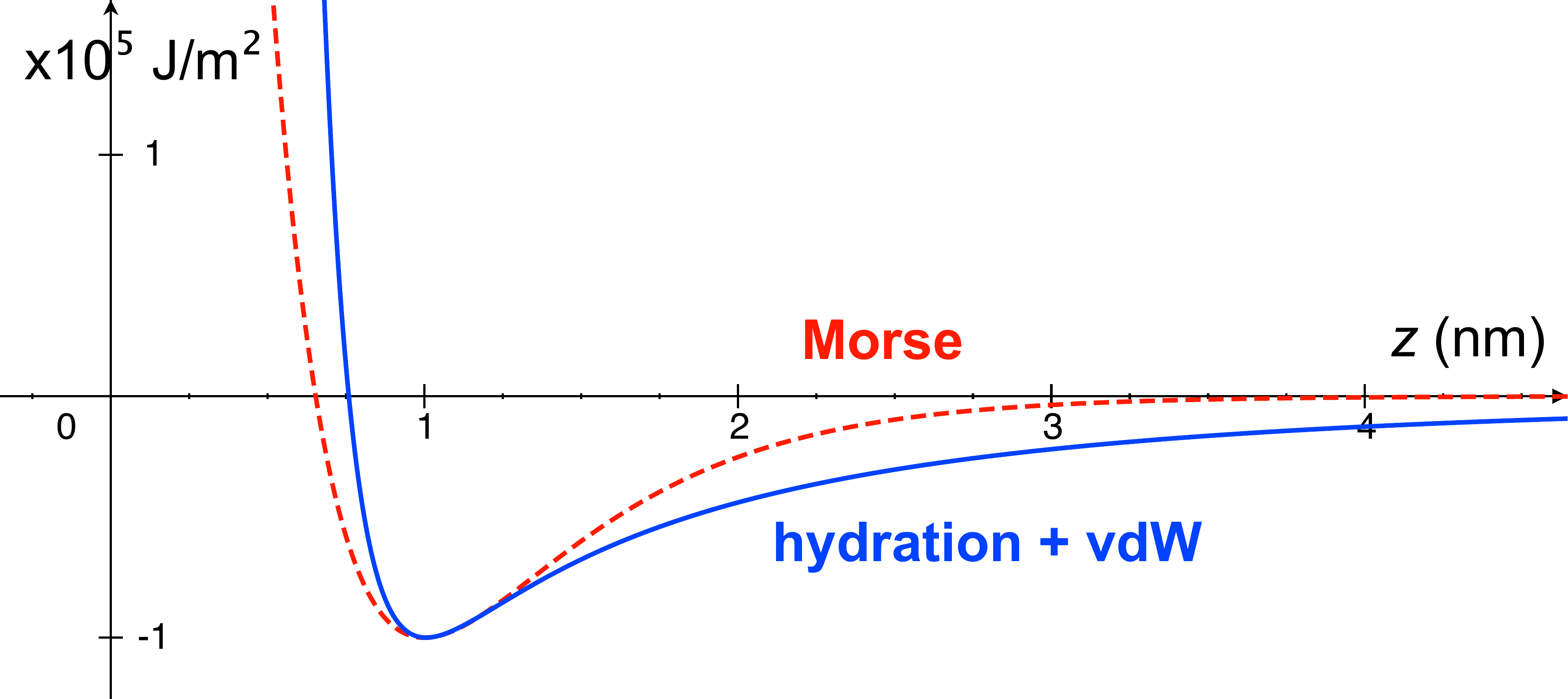}
\caption{Substrate-membrane potential as a function of the distance to the substrate: sum of van der Waals \eq{vdW} and hydration \eq{hyd} contributions (solid blue line) and Morse potential (dashed red line). Parameters for the Morse potential are adjusted to have a good match at the minimum with the true potential ($P_{\rm hyd}=1.5\times10^{-2}$~J/m$^2$, $\lambda_{\rm hyd}=0.1$~nm, $A_{\rm H}=1.5\times10^{-21}$~J, $\epsilon=3$~nm, $D=10^{-5}$~J/m$^2$, $\alpha=2$~nm$^{-1}$, and $d=1$~nm).}
\label{potential}
\end{center}
\end{figure}

In order to focus on the physical mechanisms of unbinding, we choose to mimic the adsorption potential by a simpler one, the Morse potential 
\be
V_{\rm Morse}(z)=D\left[e^{-2\alpha(z-d)}-2e^{-\alpha(z-d)}\right]
\label{Morse}
\ee
which contains the two essential features described above (repulsion at short length scales and attraction at intermediate ones, see \ref{potential}). It is characterized by only three parameters, its depth $D$, its range $\alpha$, and the position of its minimum $d$. 
In principle, the following variational treatment could be done for the potential defined in \eqs{vdW}{hyd}. However the Fourier transform of \eq{vdW}, needed in the following calculations, leads to singular parts that should be regularized using heuristic functions~\cite{podgornik}.

\textit{A priori}, replacing the potential \eq{vdW} which varies as $-z^{-3}$ for $z\gg \epsilon \simeq 3$~nm by an exponential decreasing potential \eq{Morse} could modify the nature of the transition. Indeed, we move from the weak-fluctuation regime, $|V_{\rm vdW}|\gg |V_{\rm fl}|$, to the intermediate-fluctuation regime, $|V_{\rm Morse}|\sim |V_{\rm fl}|$ (where $V_{\rm fl}$ is the fluctuation-induced repulsion which decays exponentially at large distances). However, it has been shown that in these two regimes, first-order transitions can occur~\cite{lipowskyEPL,grotehans2}, and in the intermediate-fluctuation regime their occurrence is related to the comparison of the two decay lengths (see below).

At finite temperature, thermal membrane shape fluctuations induce an entropic confinement of the bending modes (which depends on $\kappa$ and $\sigma$) and modify in turn the repulsive component of the interaction~\cite{helfrich2,evans,podgornik}. Since the full calculation of the partition function $\mathcal{Z}=\int\mathcal{D}h\,e^{-\beta\mathcal{H}[h]}$ where $\beta=(k_BT)^{-1}$, is untractable, we use in the following, a variational approach where the full Hamiltonian $\mathcal{H}$ is approximated by a trial Gaussian one 
\be
\mathcal{H}_0 = \frac12 \int_\mathcal{S} \mathrm{d}^2\br \left\{ \sigma \left(\bnabla h(\br) \right)^2 + \kappa \left(\bnabla^2 h(\br) \right)^2 +A [h(\br)-\eta]^2 \right\}
\label{Hvar}
\ee 
where the potential is harmonic with two variational parameters, the spring constant $A$, and the shifted equilibrium distance $\eta$.

The variational free energy reads in the Gibbs-Bogoliubov form
\begin{equation}
F_{\mathrm{var}}=F_0 + \langle\mathcal{H}-\mathcal{H}_0 \rangle_0,
\label{Fvar}
\end{equation}
where $\beta F_0=-\ln\mathcal{Z}_0$ is the free energy associated with the variational Hamiltonian~(\ref{Hvar}) and the subscript 0 refers to quantities calculated using \eq{Hvar}.
The Gibbs inequality ensures that $F_{\mathrm{var}}\geq F_{\mathrm{exact}}$ when $F_{\mathrm{var}}$ is minimized
with respect to the variational parameters. Their values will thus be determined by minimizing \eq{Fvar} with respect to $A$ and $\eta$. It must be emphasized, however, that we restrict our choice of variational Hamiltonians to the subclass of quadratic (and thus symmetrical) potentials, and by  doing so, we obtain an approximate value of the true minimum of the free energy or equivalently the true values of the unbinding transition temperature and of the critical exponents.

The calculations of \eq{Fvar} with the Morse potential, \eq{Morse}, have been done in Ref.~\cite{dauxois} in the context of DNA denaturation, i.e. for a one dimensional string in a space of two dimensions. In our case, the free energy $F_0$ is
\be
F_0=-\frac{k_BT\mathcal{S}}{2}\int \frac{\mathrm{d}^2\bq}{(2\pi)^2}\ln\left(\frac{2\pi k_BT}{A+\sigma q^2+\kappa q^4}\right),
\ee
and by defining $u(\br)=h(\br)-\eta$ we get
\be
\langle V_{\rm Morse}(h)\rangle_0 =  D \left[e^{2\alpha(d-\eta)}e^{2\alpha^2\langle u^2\rangle_0}-2\,e^{\alpha(d-\eta)}e^{\frac{\alpha^2}2\langle u^2\rangle_0}\right].
\ee
Minimization of $F_{\mathrm{var}}$ with respect to $\eta$ yields
\be
\eta-d = \frac32\alpha\langle u^2\rangle_0 >0
\label{etamin}
\ee
where the average mean square value of $u(\br)$ is
\bea
\langle u^2\rangle_0 &=& \frac2{\mathcal{S}}\frac{\partial F_0}{\partial A} = k_BT\int \frac{\mathrm{d}^2\bq}{(2\pi)^2}\frac1{A+\sigma q^2+\kappa q^4}\nonumber \\ &=& \frac{k_BT}{2\pi\sigma} \; g\left(\frac{4A\kappa}{\sigma^2}\right)\label{g}
\eea
and the function $g(x)=\int_0^\infty \mathrm{d}t/(x+2 t+t^2)$ is defined by~\cite{lipowsky_review}
\be
g(x) = \left\lbrace
\begin{array}{l}
\mathrm{arctanh}(\sqrt{1-x})/\sqrt{1-x}  \quad \mathrm{for}\quad x\leq1\\
\arctan(\sqrt{x-1})/\sqrt{x-1} \quad \mathrm{for}\quad x\geq1
\end{array}\right. .
\ee

In the case of vanishing surface tension $\sigma\to0$, since $g(x\to\infty)\simeq\frac{\pi}{2\sqrt{x}}$, Eq.~(\ref{g}) reduces to  the classical result of Helfrich with no surface tension~\cite{helfrich2,podgornik}
\be
A= \frac{(k_BT)^2}{(8\langle u^2\rangle_0)^2 \kappa}.
\label{Akappa0}
\ee
Helfrich closed the calculation self-consistently and found $\langle u^2\rangle_0=\mu_s d^2/4$ with $\mu_s\approx 1/6$ for a simple steric component at $z=0$~\cite{helfrich2}. In our variational approach, the solution depends on the surface tension $\sigma$ and on the respective values of the Morse  potential parameters, $D$, $\alpha$, and $d$.

Since $F_{\mathrm{var}}=F_0+\mathcal{S}\langle V\rangle_0-A\frac{\partial F_0}{\partial A}$, the minimization with respect to $A$ yields $A\frac{\partial^2 F_0}{\partial A^2}=\mathcal{S}\frac{\partial \langle V\rangle_0}{\partial A}$ as already noticed by Podgornik and Parsegian~\cite{podgornik} and we finally get an implicit equation for $A$
\be
A = 2\alpha^2 D \, e^{-\frac23\alpha (\eta-d)}
   = 2\alpha^2 D \exp\left[- \frac{k_BT \alpha^2}{2\pi\sigma} \; g\left(\frac{4A\kappa}{\sigma^2}\right)\right] 
\label{implicitA}
\ee
Solving Eq.~(\ref{implicitA}), which leads to the variational parameters $A^*$ and $\eta^*$, amounts to finding the lowest variational energy $F_{\mathrm{var}}^*=F_{\mathrm{var}}(A^*,\eta^*)$, solution of the problem. 

Let us denote by $\bar F_{\mathrm{var}}$ the variational free energy where $\eta$ is replaced by its expression in Eq.~(\ref{etamin})
\bea
\frac{\beta\bar F_{\mathrm{var}}(A)}{\mathcal{S}} &=& \frac{\sigma}{16\pi\kappa}\left[f\left(\frac{4A\kappa}{\sigma^2}\right)-\frac{4A\kappa}{\sigma^2}g\left(\frac{4A\kappa}{\sigma^2}\right)\right]\nonumber\\ &-&\beta D \exp\left[- \frac{k_BT \alpha^2}{2\pi\sigma} \; g\left(\frac{4A\kappa}{\sigma^2}\right)\right]
\label{Fvar1}
\eea
where the function $f(x)=\int_0^x g(t)\mathrm{d}t$ is
\be
f(x) = \left\lbrace
\begin{array}{l}
-2\sqrt{1-x}\,\mathrm{arctanh}(\sqrt{1-x})-\ln\left(\frac{x}4\right), \quad x\leq1\\
2\sqrt{x-1}\arctan(\sqrt{x-1}) - \ln\left(\frac{x}4\right), \quad x\geq1
\end{array}\right.
\label{f}
\ee
The free energy of an unbound membrane which fluctuates freely in the bulk is given by Eq.~(\ref{Fvar1}) with $A=0$. Thus, the membrane remains weakly adsorbed on the substrate as long as $\bar F_{\mathrm{var}}(A)<\bar F_{\mathrm{var}}(0)=0$.
We introduce the renormalized parameters
\be
\tilde A=\frac{4A\kappa}{\sigma^2}, \qquad \Xi=  \frac{8\alpha^2 \kappa D}{\sigma^2}, \qquad \theta=\frac{2\pi\sigma}{k_B\alpha^2},
\ee
where the coupling parameter $\Xi$ is simply the rescaled second derivative of the Morse potential at the minimum, and $\theta$ is a temperature scale. Since thermal fluctuations decrease the interaction between the membrane and the substrate, we expect that for a finite temperature, we have $A(T)\leq 2\alpha^2 D$ or $\tilde A(T)\leq \Xi$.
 
From \eq{etamin}, one sees that the average height of the bilayer, $\eta$, is proportional to the average mean square of height fluctuations $\langle u^2\rangle_0$. Since $g(x)$ is a monotonic decreasing function which tends to 0 for large $x$ and diverges for $x=0$, decreasing $\tilde A$, i.e. decreasing $A$ or $\kappa$ or alternatively increasing $\sigma$, destabilizes the supported membrane. This increase of $\eta$ has been recently observed by heating double supported bilayers which makes $\kappa$ decreasing ($\sim 200 k_BT$ in the gel phase to $1-3 k_BT$ at its minimum)~\cite{lecuyer}.

It is interesting to note that naively one would expect the membrane to desorb when the fluctuation contribution in \eq{etamin} diverges, i.e. for $\tilde A\to0$. Technically speaking, this type of transition would be continuous. 
However, as said in the Introduction, discontinuous transitions can occur. Indeed, close to the transition we have $\tilde A\ll1$ and \eqs{etamin}{g} yields $\eta-d\simeq -3/(8\pi) k_BT\alpha/\sigma \ln(\tilde A/4)$. The variational free energy  \eq{Fvar1}  can be written as 
\bea
\frac{\beta F_{\mathrm{var}}(\eta)}{\mathcal{S}} &\simeq& \frac{\sigma}{16\pi\kappa}\left\{2\exp\left[- \frac23 \frac{2T}{\theta}\alpha (\eta-d)\right] \right.\nonumber\\
&-& \left.\frac{\Xi\theta}{T} \exp\left[- \frac23 \alpha (\eta-d)\right]\right\}
\label{Fvar*}
\eea
and thus comparing both exponentials and prefactors, we find an unbinding transition in two cases: i) a continuous transition for $T/\theta\leq2$ and $\Xi\leq2T/\theta\leq \Xi_c=4$, and ii) a discontinuous transition for $T/\theta>2$ and large $\Xi>2T/\theta\geq\Xi_c=4$.

\begin{figure}[t]
\begin{center}
(a)\includegraphics[width=7cm]{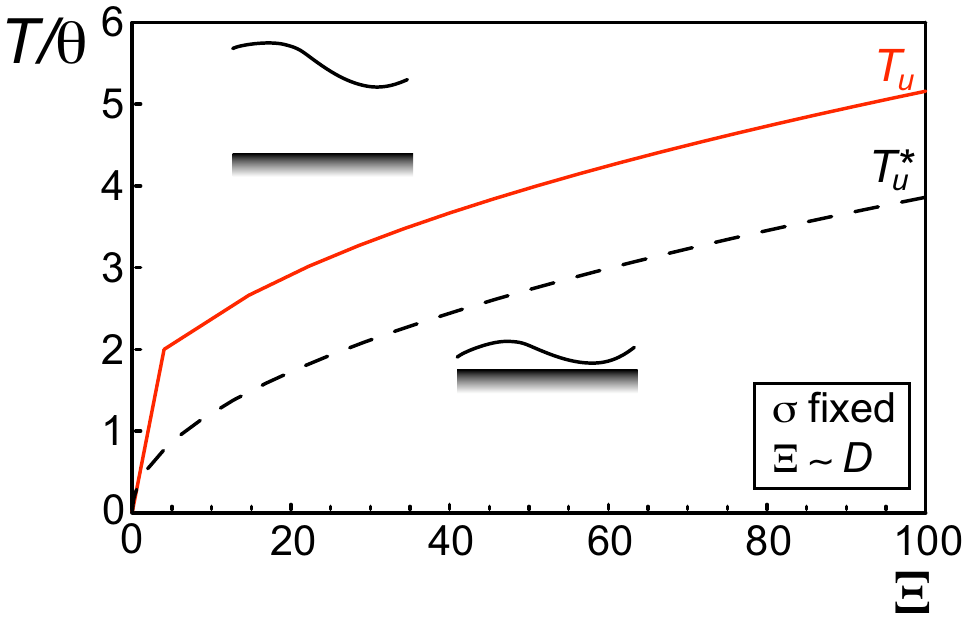}\hspace{1cm}
(b)\includegraphics[width=7cm]{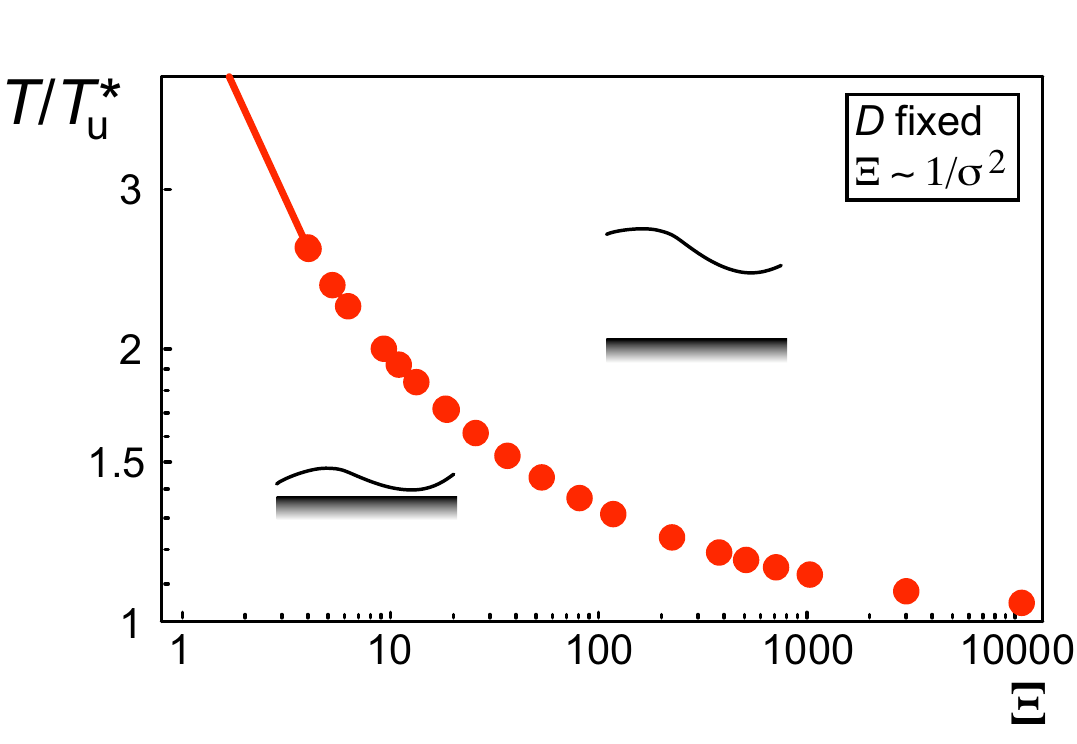}
\caption{(a) Phase diagram of an adsorbed membrane in a Morse potential for a fixed value of the surface tension $\sigma$. The solid (red) line corresponds to the rescaled unbinding temperature $T_u/\theta$ vs. the coupling parameter $\Xi$ which measures the bare adsorption strength. It divides the phase diagram in two parts, where the membrane is bound (bottom) and unbound (top). The dashed line corresponds to the system with vanishing surface tension given by \eq{signull}. (b) Same phase diagram in the $(\Xi,T/T_u^*)$ space, i.e. for constant adsorption strength, $D$, and varying $\sigma$. The surface tension increases the unbinding temperature (dots) which behaves like $1/\sqrt{\Xi}$ for $0\leq\Xi\leq4$ (solid line) and reaches the asymptote 1 for large $\Xi$.}
\label{1memb}
\end{center}
\end{figure}

More quantitatively, the unbinding temperature $T_u$ is defined by  $\bar F_{\mathrm{var}}(A)=0$ together with Eq.~(\ref{implicitA}), which yields
\bea
\frac{\theta}{T_u} &=& \frac{f(\tilde A)}{\tilde A}-g(\tilde A)\label{fnull}\\
\tilde A &=& \Xi \exp\left[- \frac{T_u}{\theta} \; g(\tilde A)\right]\label{Dfnull}
\eea
The solution of this system is found numerically and the phase diagram is shown in \ref{1memb}(a) keeping $\sigma$ fixed. At low temperature, the membrane is bound to the substrate and the binding region grows with the adsorption strength $\Xi$, which is in this case proportional to $D$.

In the vanishing surface tension case, $\sigma\to0$, one finds, from \eqs{fnull}{Dfnull} and \eq{Akappa0}, that the unbinding temperature is given by $T_u(\sigma\to0)/\theta = 2/(\pi\sqrt{e})\sqrt{\Xi}$, i.e.
\be
T_u^*\equiv T_u(\sigma=0)= 8\sqrt{\frac2{e}}\frac{\sqrt{\kappa D}}{\alpha k_B}.
\label{signull}
\ee
We obtain the correct scaling for $T_u$~\cite{lipowsky_review}. \eq{signull}, in unrescaled units, corresponds to the dashed line in \ref{1memb}(a).
Note that with the realistic parameter values given in \ref{potential}, we find $T_u^*=0.94\, T_{\rm room}$.

The same phase diagram is re-plotted in \ref{1memb}(b) using the parameters $T/T_u^*$ and $\Xi$. The adsorption strength $D$ is kept fixed, and thus the  parameter $\Xi$ now controls the surface tension. One observes that the unbinding temperature increases when $\Xi$ decreases, hence when the surface tension increases. Indeed, the system is entropically stabilized when $\sigma\neq0$, since more degrees of freedom are accessible to the bound membrane at a given temperature. At low $\Xi$, the unbinding temperature follows the limiting law $T_u/T_u^*\simeq \frac{\pi\sqrt{e}}{\sqrt{\Xi}}$. Indeed Eq.~(\ref{fnull}) leads to $T_u\to 2\theta$ for $\tilde A\to 0$. This limiting form corresponds to the solid line in \ref{1memb}(b). At large $\Xi$, i.e. for $\sigma\to0$, $T_u/T_u^*\to1$.

\begin{figure}[t]
\begin{center}
(a)\includegraphics[width=7.5cm]{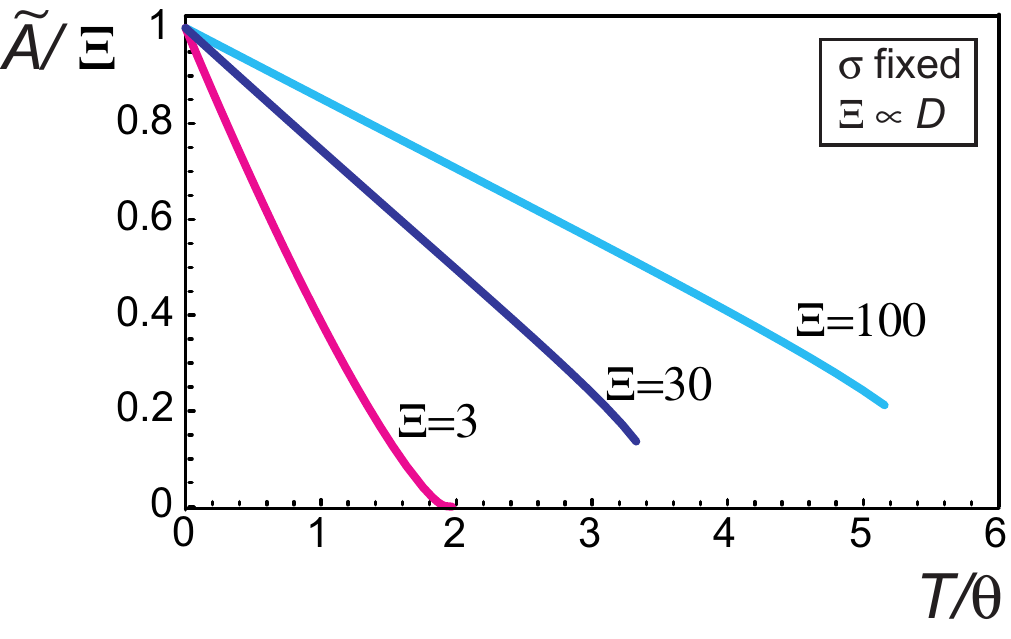}
(b)\includegraphics[width=7cm]{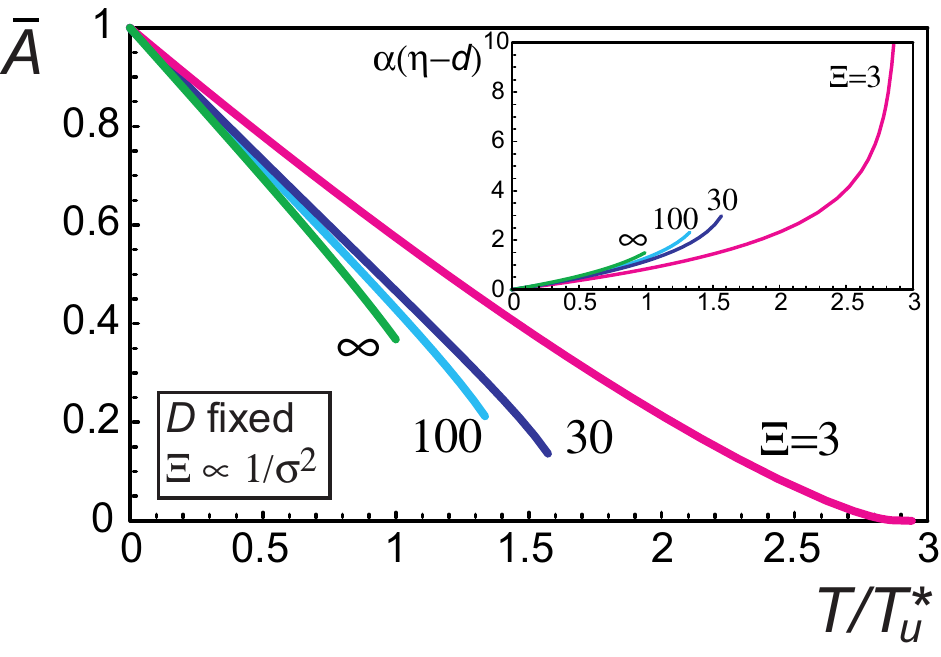}
\caption{(a) Variation of the variational parameter $\tilde A$ rescaled by the adsorption strength $\Xi$ as a function of rescaled temperature $T/\theta$ ($\sigma$ constant) for different values of $\Xi=3,30,100$. The parameter $\Xi$ is in this case proportional to the adsorption strength $D$. The fluctuation induced repulsion increases when $T$ increases thus leading to a decrease of $\tilde A$, until we reach the critical point at $T=T_u$. (b) Same as (a) with the temperature rescaled by the unbinding temperature at $\sigma=0$. $\Xi$ is thus proportional to $\sigma^{-2}$ and the case $\Xi\to\infty$ corresponds to vanishing surface tension, Eq.~(\ref{nost}). The associated average distance between the substrate and the membrane, $\eta$, given by \eq{implicitA} is plotted in the inset.}
\label{1membplus}
\end{center}
\end{figure}

The disjoining pressure, or mechanical stress~\cite{evans}, is 
\be
p(T)=-\frac1{\mathcal{S}}\frac{\partial F_{\mathrm{var}}}{\partial d}=\frac{A(T)}{3\alpha},
\ee
since $\partial F_{\mathrm{var}}/\partial d=\partial \langle V_{\rm Morse}(h)\rangle_0/\partial d$. Hence the variational parameter $A(T)$, plotted in \ref{1membplus} for various values of $\Xi$, readily gives the disjoining pressure as a function of temperature, which can be measured experimentally.
To compare this case to the vanishing surface tension case, we rescale the parameter $A$ by the spring constant in the Morse potential $2\alpha^2D$ as $\bar A=A/(2\alpha^2 D)=\tilde A/\Xi$. Equation~(\ref{Dfnull}) becomes
\be
\bar A=\frac{A}{2\alpha^2 D}=\exp\left(-\frac2{\pi\sqrt{e}} \frac{T}{T_u^*} \sqrt{\Xi} \, g(\Xi\bar A) \right).
\label{nost}
\ee
Note however, that our variational approach, which belongs to the ``superposition of fluctuation-induced and direct forces'' type of approaches, is not appropriate in this strong-fluctuating regime (with $\sigma=0$ and thus $|V_{\rm fl}|\gg|V_{\rm Morse}|$), since, in this regime, only renormalization group approaches yields the correct (second) order of the transition~\cite{lipowsky_review}.

In \ref{1membplus}(a) and (b) are plotted $\tilde A$ [\eq{Dfnull}] and $\bar A$ [\eq{nost}], as a function of $T$ at constant $\sigma$ and constant $D$, respectively. First of all, one observes that $A$ (and $\eta$) depend on $T$ which is a signature of the non-linear potential as for thermal expansion in solids. We recover the straight result that at $T\to0$, $A(T=0)=V_{\rm Morse}''(d)=2\alpha^2 D$, i.e. the spring constant with no entropic induced repulsion [\ref{1membplus}(b)]. Furthermore, at fixed $T$, $\tilde A$ increases with $\Xi$ (or $D$) faster than $\Xi$, and $\bar A$ decreases slightly when $\Xi$ increases (or $\sigma$ decreases). This is an important result in the experimental context, since the surface tension of supported membrane may vary due to defects where the membrane is pinned or depending on the experimental protocol (e.g. washing processes or superficial pressure exerted at the edge). The variational mean height  of the bilayer $\eta$, which is the true order parameter of the transition, is deduced following Eq.~(\ref{implicitA}). On observes that, for $\Xi=3$, $A=0$ at the unbinding temperature (or $\eta$ diverges). More generally, as suggested in \eq{Fvar*}, our approach yields a continuous transition (or second order) for $\Xi\leq 4$ (solid line in the constant $D$ phase diagram shown in \ref{1memb}(b)). For $\Xi>4$, $A(T_u)$ is finite which makes the transition discontinuous (first order). The point ($\Xi_c=4,T_c/\theta=2$) is thus a tricritical Lifshitz point, the surface tension $\sigma$ controlling the order of the transition.


\section{Two solid-supported membranes}


In this section we consider two membranes superimposed on a substrate, with an adsorption potential as \eq{Morse} for the first membrane and a similar one for  the inter-membrane potential. Since the range of the adsorption potential is small ($\alpha^{-1}\sim d$, a few nm), the second stacked bilayer does not feel directly the substrate and is adsorbed only through the inter-membrane potential. We model these two potentials as Morse potentials $V_i$ defined in \eq{Morse} with parameters $D_i$, $d_i$ and $\alpha_i$ where $i=1$ refers to the lower membrane and $i=2$ to the upper one. The Hamiltonian of the system is
\bea
\mathcal{H} &=& \int_\mathcal{S} \mathrm{d}^2\br \sum_{i=1}^2\left\{ \frac12\left[\sigma \left(\bnabla h_i(\br) \right)^2 + \kappa \left(\bnabla^2 h_i (\br) \right)^2\right]\right.\nonumber\\ &&+ \left. V_i[h_i(\br)-h_{i-1}(\br)] \right\}
\label{Hhelfrich2}
\eea
where $h_0=0$ is the position of the substrate plane. The variational Hamiltonian reads
\bea
\mathcal{H}_0 &=& \frac12 \int_\mathcal{S} \mathrm{d}^2\br \sum_{i=1}^2 \left\{ \sigma \left(\bnabla h_i(\br) \right)^2 + \kappa \left(\bnabla^2 h_i (\br) \right)^2 \right.\nonumber\\ && \left.+A_i [h_i(\br)-h_{i-1}(\br)-\eta_i]^2 \right\}\nonumber\\
&=& \frac12\int_\mathcal{S} \mathrm{d}^2\br \,\bu^T(\br)\,{\bf U}\,\bu(\br)
\label{Hvar2}
\eea
with  $\bu^T(\br)=(h_1(\br)-\eta_1,h_2(\br)-\eta_2-\eta_1)$ and
\be
{\bf U} =(- \sigma\nabla^2 + \kappa\nabla^4)\mathbf{I_2} +\left(\begin{array}{cc}
 A_1 + A_2 &  - A_2 \\
-A_2  &  A_2 
\end{array}\right)
\label{defU}
\ee
where $\mathbf{I_2}$ is the identity matrix. In the eigenbasis the modes are decoupled with eigenvalues (of the last matrix in \eq{defU})
\be
\lambda_{1,2} = \frac{A_1}2+A_2\pm \sqrt{\frac{A_1^2}4+A_2^2}
\label{lambda12}
\ee
and the Hamiltonian \eq{Hvar2} becomes diagonal. The normal modes $\bv^T(\br)=(v_1(\br),v_2(\br))$ are defined through $\bu(\br)={\bf P}\, \bv (\br)$ where ${\bf P}$ is a matrix of rotation of angle $\phi \in [0,\frac\pi2]$ defined by
\be
\tan\phi=\frac{A_2}{\sqrt{\frac{A_1^2}4+A_2^2}-\frac{A_1}2}.
\label{theta}
\ee
One observes that $\lambda_1>\lambda_2\geq0$ where $\lambda_2=0$ for $A_1=0$ or $A_2=0$. The case $A_1=0$ ($\phi=\pi/4$) corresponds to a free stack of two membranes remaining bound together, whereas $A_2=0$ ($\phi=0$) corresponds to the case of a second membrane free while the first one remains bound. 

By proceeding as above, the free energy $F_0$ is simply given by the sum over the modes
\bea
F_0 &=& -\frac{k_BT\mathcal{S}}{2}\sum_{i=1}^2 \int \frac{\mathrm{d}^2\bq}{(2\pi)^2}\ln\left(\frac{2\pi k_BT}{\lambda_i+\sigma q^2+\kappa q^4}\right)\nonumber\\
&=& \frac{k_BT\mathcal{S}\sigma}{16\pi\kappa}\left[f\left(\frac{4\lambda_1\kappa}{\sigma^2}\right)+f\left(\frac{4\lambda_2\kappa}{\sigma^2}\right)\right].
\eea
where $f(x)$ is defined in \eq{f}. Minimization of the variational free energy with respect to the four variational parameters $A_i$ and $\eta_i$, yields the generalization of Eqs.~(\ref{etamin}) and~(\ref{implicitA})
\bea
\eta_i-d_i &=& \frac32\alpha_i\langle  (u_i-u_{i-1})^2\rangle_0\\
A_i &=& 2\alpha_i^2 D_i \exp\left[-\alpha_i^2 \; \langle (u_i-u_{i-1})^2\rangle_0\right]
\label{Ai}
\eea
where
\be
\langle (u_i-u_{i-1})^2\rangle_0=\frac{k_BT}{2\pi\sigma} \sum_{j=1}^2 g\left(\frac{4\lambda_j\kappa}{\sigma^2}\right)\,\frac{\partial \lambda_j}{\partial A_i}.
\label{ui}
\ee

By replacing the $\eta_i$ as a function of $A_i$ and writing with a tilde dimensionless quantities, $\tilde A _i=4A_i\kappa/\sigma^2$ and  $\tilde \lambda _i= 4\lambda_i\kappa/\sigma^2$, the variational free energy $\bar F_{\mathrm{var}}(A_1,A_2)$, corresponding to \eq{Fvar1} for a single supported membrane, is
\bea
\frac{\beta\bar F_{\mathrm{var}}}{\mathcal{S}} &=& \frac{\sigma}{16\pi\kappa}\left[f(\tilde \lambda_1)+f(\tilde \lambda_2)- \sum_{i,j=1}^2\ \tilde \lambda_i g(\tilde \lambda_j)\,\frac{\partial \tilde\lambda_j}{\partial \tilde A_i}\right] \nonumber\\&&- \sum_{i=1}^2\beta D_i \exp\left[- \frac{k_BT \alpha_i^2}{2\pi\sigma} \sum_{j=1}^2 g(\tilde \lambda_j)\,\frac{\partial \tilde \lambda_j}{\partial \tilde A_i}\right]
\label{Fvar2}
\eea
where $\tilde\lambda_{1,2}(\tilde A_1,\tilde A_2)$ are given in Eq.~(\ref{lambda12}).

The unbinding of the adsorbed stack can \textit{a priori} follow three different scenarii~: i) the upper membrane desorbs and diffuses freely in the bulk while the lower one remains adsorbed ($A_2=0$); ii) the stack as a whole evolves freely in the bulk ($A_1=0$); or iii) the two membranes are completely de-stacked and desorbed. Hence, to study quantitatively this unbinding transition, the variational free energy, Eq.~(\ref{Fvar2}), should be compared to the variational free energies of the system in the three configurations described above
\be
\Delta \bar F_{\mathrm{var}}(A_1,A_2)= \bar F_{\mathrm{var}}(A_1,A_2)-\bar F_{\mathrm{var}}
\ee
where $\bar F_{\mathrm{var}}$ is the variational free energy of the (possibly partially) unbound system. Since the variational free energy of a freely fluctuating membrane in the bulk has been chosen as the reference of energy, these free energies, $\bar F_{\mathrm{var}}$, are respectively i)  Eq.~(\ref{Fvar1}) at its minimum, $\bar F_{\mathrm{var}}(\tilde A_1^*)$; ii) Eq.~(\ref{Fvar2}) with $\tilde\lambda_1=2\tilde A_2$ and $\tilde\lambda_2=0$ at its minimum $\tilde A_2^*$
\bea
\frac{\beta\bar F_{\mathrm{var}}(\tilde A_2^*) }{\mathcal{S}} &=& \frac{\sigma}{16\pi\kappa}\left[f(2\tilde A_2^*)-2\tilde A_2^*\,g(2\tilde A_2^*)\right]\nonumber\\&&-\beta D_2\exp\left[- \frac{k_BT \alpha_2^2}{\pi\sigma}  g(2\tilde A_2^*)\right]
\eea
and iii) $\bar F_{\mathrm{var}}(0)=0$.

\begin{figure}[t]
\begin{center}
(a)\includegraphics[width=7.5cm]{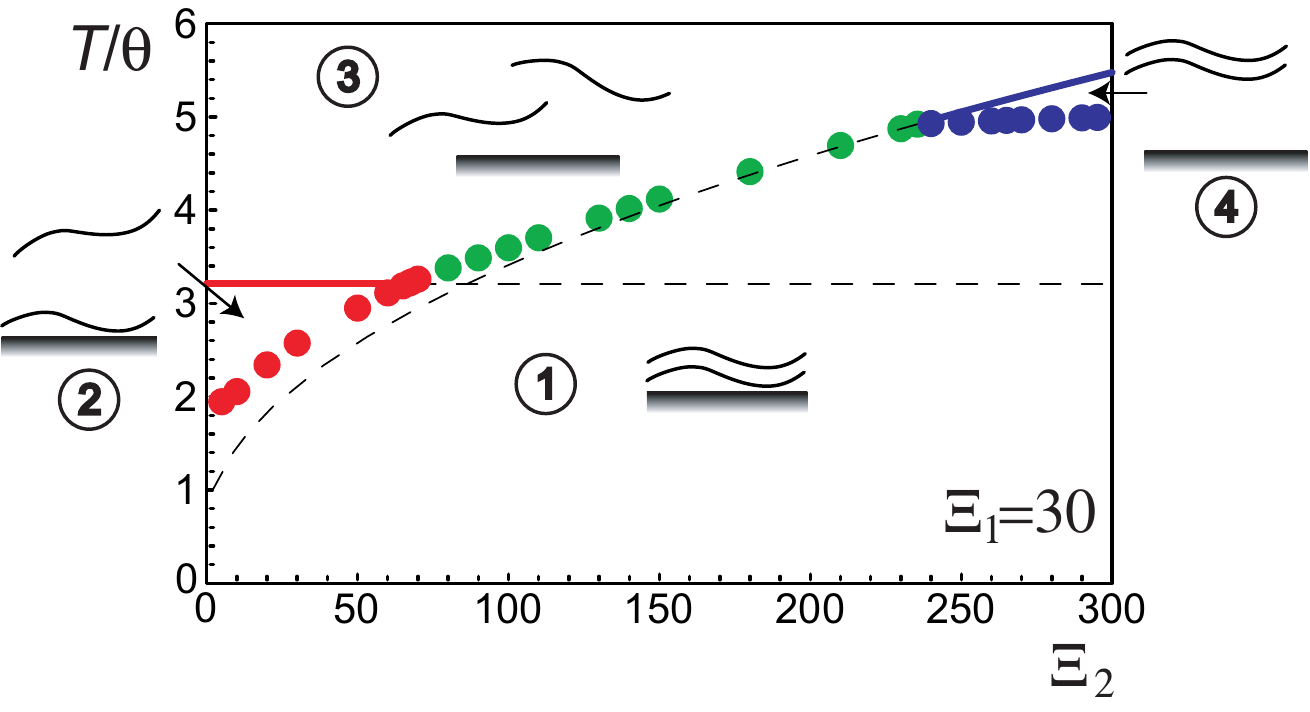}
(b)\includegraphics[width=7.5cm]{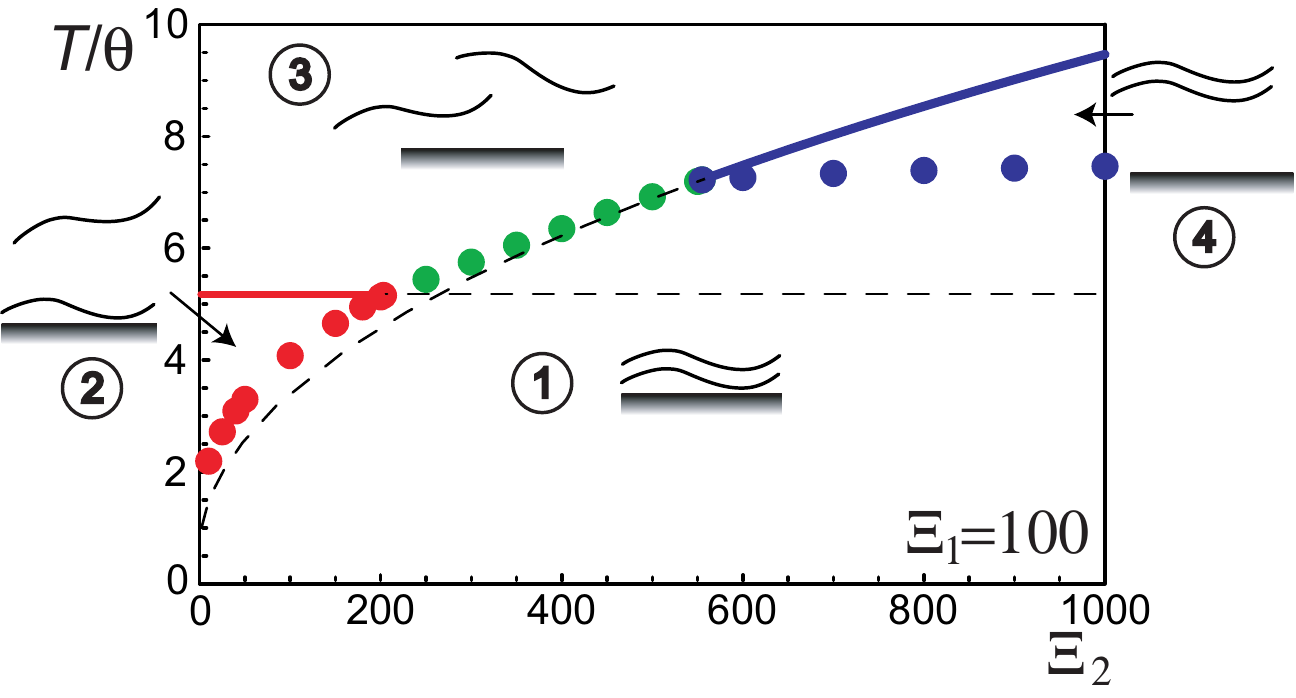}
\caption{Phase diagram of two adsorbed membranes in a Morse potential, (a) $\Xi_1=30$; (b) $\Xi_1=100$. The colored dots corresponds to the rescaled unbinding temperature $T_u/\theta$ of the adsorbed stack vs. the coupling parameter $\Xi_2$ which measures the strength of the inter-membrane potential. Solid lines (dashed for unstable case) correspond to the unbinding of the single membrane when the upper is unbound (red) and of the two membranes in the free stack (blue). The diagram is thus divided in four regions, where the stack is bound (region~1), the upper membrane desorbs (2), both membranes are unbound~(3), and the stack is unbound (4).}
\label{2memb}
\end{center}
\end{figure}
Hence, the unbinding temperature, $T_u$, is solution of Eq.~(\ref{Ai}) and $\Delta \bar F_{\mathrm{var}}(A_1,A_2)=0$, i.e. the values of $A_1$, $A_2$ and $T_u$ are solutions of the three equations:
\bea
&&\sum_{i=1}^2 \left[f(\tilde\lambda_i)-\sum_{j=1}^2 g(\tilde\lambda_j) \tilde A_i\,\frac{\partial \tilde\lambda_j}{\partial \tilde A_i} - \frac{\theta}{T_u} \tilde A_i\right] \nonumber\\ &=& \left\{ 
\begin{array}{l} 
f(\tilde A_1^*) - g(\tilde A_1^*)\tilde A_1^*- \frac{\theta}{T_u}\tilde A_1^* \\
f(\tilde 2A_2^*) - 2g(\tilde 2 A_2^*)\tilde A_2^*- \frac{\theta}{T_u}\tilde A_2^*\\
0
\end{array}\right.\label{var1}
\\
\tilde A_i &=& \Xi_i \exp\left[- \frac{T_u}{\theta} \sum_{j=1}^2 g(\tilde \lambda_j)\,\frac{\partial \tilde\lambda_j}{\partial \tilde A_i} \right]\; \; i\in\{1,2\}.
\label{var2}
\eea
The phase diagram is shown in \ref{2memb}(a) for $\Xi_1=30$ and in \ref{2memb}(b) for $\Xi_1=100$ ($\alpha_1=\alpha_2$), and $\Xi_2/\Xi_1$ varying between 0 and 10. The surface tension $\sigma$ and bending modulus $\kappa$ of both membranes are kept fixed, which sets the parameters $\Xi_i$ to be proportional to the two adsorption strengths $D_i$. Four distinct regions appear, related to the four free energies defined above. 

\begin{figure*}[t]
\begin{center}
\includegraphics[width=16cm]{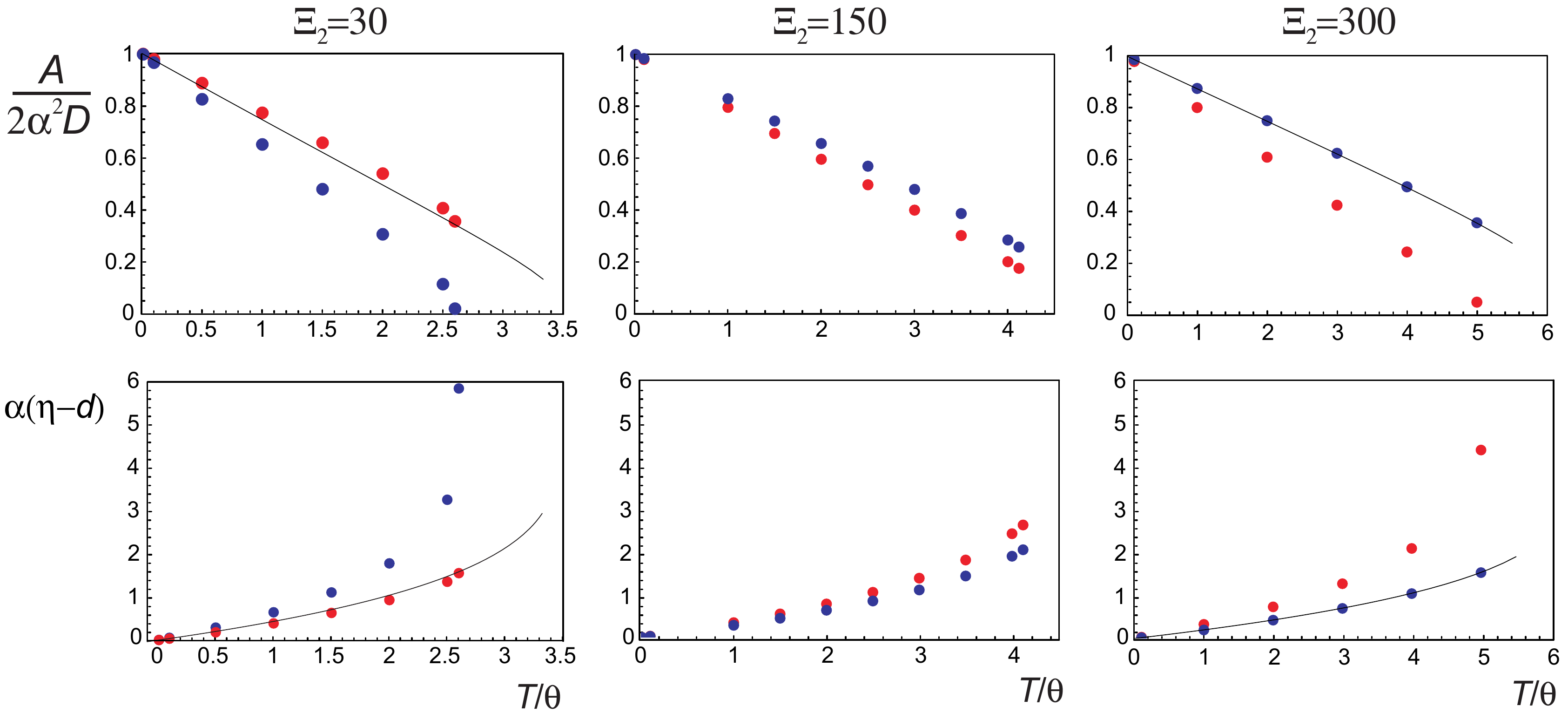}
\caption{Top: Variational parameter $\tilde A/\Xi=A/(2\alpha^2D)$ vs temperature $T/\theta$ for $\Xi_2=30,150, 300$ ($\Xi_1=30$). Red dots correspond to the lower bilayer, blue ones to the upper one. For comparison sake, solid lines correspond to one supported membrane for $\Xi=30$ and a free stack of two bilayers for $\Xi_2=300$. Bottom: Corresponding average distance between membranes, $\alpha(\eta-d)$, given by \eq{implicitA}, vs. $T/\theta$.}
\label{figlambda}
\end{center}
\end{figure*}

For low $\Xi_2<2\Xi_1$ and by increasing temperature, the stack is progressively peeled up, the upper bilayer unbinding at a temperature lower than for the bilayer close to the substrate (region~2). However, for intermediate values of $\Xi_2$ [$2\Xi_1<\Xi_2<235.5$ in \ref{2memb}(a) and 555 in \ref{2memb}(b)], the stack unbinds completely at the transition, defining one unique unbinding temperature $T_u(\Xi_2)$. One then enters in region~3 where both membranes are unbound. This unbinding transition temperature of the adsorbed stack is larger than for a single adsorbed membrane. Indeed, the upper membrane reinforces the adsorption of the lower one by ``squeezing'' it to the substrate. Moreover, the region in the phase diagram where the stack exists is larger than when there is no adsorbing surface. This is due to the slight decrease of the first membrane height fluctuations induced by the presence of the wall.  
Finally, for large enough $\Xi_2$ values, a fourth region appears (region~4) where the stack unbinds as a whole, the two membranes remaining bound together because of their strong mutual attraction.

In \ref{figlambda} are plotted the variational parameters as a function of $T/\theta$, for three different values of $\Xi_2/\Xi_1= 1, 5$ and 10 ($\Xi_1=30$), corresponding to the three regimes just described and shown in \ref{2memb}(a).
Clearly, one observes that for the upper membrane, $A_2(T)$ increases when $\Xi_2/\Xi_1$ increases, as expected. For $\Xi_2/\Xi_1=1$,  $A_1(T)>A_2(T)$ which is signature of the increase of height fluctuations: the upper membrane fluctuates more than the lower thanks to the fluctuations of the lower membranes which add up with its own ones. For $\Xi_2/\Xi_1=5$ we have $A_1(T)\simeq A_2(T)$, which are both larger than above, and the associated interlayer distances are almost the same. This is the reason why the stack unbinds completely at the transition. Finally, for $\Xi_2/\Xi_1=10$, $A_1(T)< A_2(T)$ and the transition of the stack is reached when $A_1(T_u)=0$.  We see that the adsorption strength of the substrate $\Xi_1$ compared to the interlayer potential tuned by $\Xi_2$ is central in computing membrane fluctuations.

More interestingly is the nature of the transition as a function of $\Xi_2$: for low $\Xi_2<2\Xi_1$ and very large $\Xi_2$ the transition is continuous occurring respectively for $A_2(T_u)=0$ and $A_1(T_u)=0$, whereas for intermediate values of $\Xi_2$ corresponding to the full unbinding of the stack (from region~1 to region~3) the transition occurs for finite values of $A_i$, i.e. it is discontinuous. This can be related to the unbinding of one bilayer with a direct interaction including a potential barrier~\cite{lipowsky_review,grotehans} which exhibits  discontinuous (or first order) unbinding transitions. In our case, the upper membrane induces such a potential barrier felt by the ``squeezed'' membrane.

\begin{figure*}[t]
\begin{center}
\includegraphics[width=16cm]{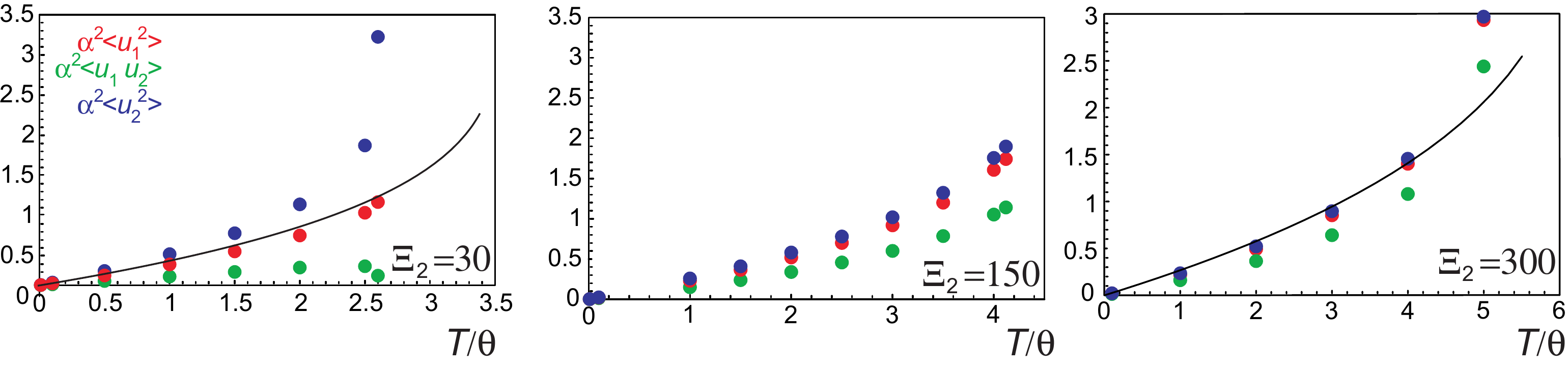}
\caption{Height-height correlation functions given by Eqs.~(\ref{u1}), (\ref{u1u2}), and (\ref{u2}) ($\alpha=\alpha_1=\alpha_2$) as a function of temperature for increasing $\Xi_2$ (same values as in \ref{figlambda}). Solid lines correspond to one supported membrane ($\Xi=30$) or a free stack of two bilayers ($\Xi_2=300$).}
\label{hhcorr}
\end{center}
\end{figure*}

Once the variational parameters, $A_i$, are determined, one has access to the fluctuations of the two membranes and the height-height correlation functions. The variational Hamiltonian being Gaussian, the structure factor is
\be
\langle v_i(\bq)v_j(\bq')\rangle=(2\pi)^2\delta_{ij}\delta(\bq+\bq')\frac{k_BT}{\lambda_i+\sigma q^2+\kappa q^4}
\label{correlations}
\ee
and the three height-height correlation functions $C_{ij}(\br)= \langle u_i(\br) u_j({\bf 0})\rangle$ are given in Appendix~\ref{correlation_functions}.
In particular, we find
\bea
\langle u_1^2\rangle = \frac{k_BT}{2\pi\sigma}\left[\cos^2\phi \, g\left(\frac{4\lambda_1\kappa}{\sigma^2}\right)+\sin^2\phi  \, g\left(\frac{4\lambda_2\kappa}{\sigma^2}\right) \right]&&\label{u1}\\
\langle u_1 u_2 \rangle = \frac{k_BT}{2\pi\sigma}\cos\phi\sin\phi\left[ g\left(\frac{4\lambda_2\kappa}{\sigma^2}\right)-g\left(\frac{4\lambda_1\kappa}{\sigma^2}\right) \right]&&\label{u1u2}\\
\langle u_2^2 \rangle = \frac{k_BT}{2\pi\sigma}\left[\sin^2\phi  \, g\left(\frac{4\lambda_1\kappa}{\sigma^2}\right) +\cos^2\phi  \,  g\left(\frac{4\lambda_2\kappa}{\sigma^2}\right)  \right]&&\label{u2}
\label{correlations3}
\eea
which are plotted in \ref{hhcorr}. Note that $\langle u_1^2\rangle=\frac{2(\eta_1-d)}{3\alpha_1}$ and $\langle (u_2-u_1)^2\rangle=\frac{2(\eta_2-d)}{3\alpha_2}$ [see \eq{ui}] are already plotted in \ref{figlambda}.
For $\Xi_2=\Xi_1$ the fluctuations of the upper membrane increase from one to two supported bilayers.  In this case, fluctuations of the lower membrane add up to the upper-membrane ones leading to a substantial increase for $T>2\theta$. Membrane 1 is thus very slightly perturbed by the presence of membrane 2 (compare lines to dots in ~\ref{figlambda} and ~\ref{hhcorr}). Finally, the correlations between both membranes are very low. When $\Xi_2$ increases, correlations between both membranes increase and their correlation functions are almost identical (for $\Xi_2/\Xi_1=5$ and 10).

In principle, the same calculations can be done for $n$ membranes (see Appendix~\ref{n_membranes}) with eigenvalues $\lambda_i$ where $i\in\{0,\dots,n-1\}$. The variational equations are the same as \eqs{var1}{var2}, where the rhs. of \eq{var1} now contains all the free energies for free bundles made of $j$ membranes ($A_{n-j+1}=0$) and adsorbed bundles made of $n-j$ membranes.
Eigenvalues and eigenmodes in the simplest case where all the variational parameters are equal, are given in Appendix~\ref{n_membranes}.

In this section, we have implicitly assumed $\alpha_2=\alpha_1$. However, one might expect to have actually different potential ranges $\alpha_2>\alpha_1$ since the inter-membrane potential is at large distances in $z^{-4}$ instead of $z^{-3}$~\cite{lipowsky_review}. It will introduce  two temperatures $\theta_1>\theta_2$ and the results will remain qualitatively the same.


\section{Dynamics of two supported lipid bilayers}


Dynamics of supported membranes are governed by Langevin equations for height displacements of membranes 1 and 2, $h_1(\br)$ and  $h_2(\br)$, written in $\bq$-space
\be
\frac{\partial h_i(\bq,t)}{\partial t} = -\Lambda_{ij} (\bq) \frac{\delta \mathcal{H}_0}{\delta h_j(-\bq,t)} + \zeta_i(\bq,t)
\label{membrane_dynamics}
\ee
where $\zeta_i(\bq,t)$ is the Fourier transform of the random noise obeying the fluctuation-dissipation theorem: $\langle  \zeta_i(\bq,t)  \zeta_j(\bq',t) \rangle =  2(2\pi^2) k_BT \Lambda_{ij}(\bq) \delta(t-t')\delta(\bq+\bq')$. The damping matrix ${\bf \Lambda}(\bq)$ takes  hydrodynamic interactions into account, both along the membrane and between adjacent surfaces (substrate--membrane 1 or membrane 1--membrane 2).

It is known that for a single membrane in an infinite liquid, the mobility is simply $\Lambda(q)=(4\mu q)^{-1} $ and \eq{membrane_dynamics} leads directly to the damping rate~\cite{seifert}
\be
\gamma_0(q)=(\kappa q^3+\sigma q)/4\mu.
\label{damping_single}
\ee
It is the ratio of the energy driving fluctuations and the viscous damping. However, in the case considered in \ref{sketch}, we expect that both the presence of the second membrane and the solid wall, which imposes a no-slip condition, will substantially modify \eq{damping_single}. In the following, we compute  $ \mbox{\boldmath${\Lambda}$}(\bq)$ in a simpler geometry, also shown in \ref{sketch}  (dashed lines), where the membranes are supposed planar, located at $z=\eta_1$ and $z=\eta_2$. 

\subsection{Hydrodynamics of two supported membranes}

The velocity flow field ${\bf v}(\bR)$ and pressure field $p(\bR)$ in the bulk, are found using the Stokes equation for an incompressible fluid of viscosity $\mu$
\bea
\nabla\cdot\bv &=& 0 \label{incompressible}\\
\mu\nabla^2\bv-\nabla p &=& 0 \label{stokes}
\eea
with the following conditions at the boundaries
\bea
p(\br,z\to\infty) &=& p_0 \qquad \forall \br\\
\bv(\br,z\to\infty) &=& 0 \qquad \forall \br\\
\bv(\br,0) &=& 0 \qquad \forall \br \label{noslip}\\
\bv(\br,\eta_{1,2}^-) &=& \bv(\br,\eta_{1,2}^+)\label{continuity}\\
\nabla_\parallel\cdot\bv_\parallel(\br,\eta_{1,2}) &=& 0  \label{incomp_memb}
\eea
\eq{noslip} imposes the no-slip condition at the substrate, and \eq{incomp_memb} ensures the membrane incompressibility.
The two fluctuating membranes impose normal forces $f_{1,2}$ at $z=\eta_{1,2}$ which are balanced by fluid stress jumps
\be
f_{1,2}=-\sigma_{zz}(\br,\eta_{1,2}^+)+\sigma_{zz}(\br,\eta_{1,2}^-)
\ee
where the stress tensor is $\sigma_{ij}=-p\delta_{ij}+\mu(\partial_j v_i+\partial_i v_j)$. From \eqs{incompressible}{incomp_memb}, one finds easily that 
\be
f_{1,2}=\delta p(\br,\eta_{1,2}).\label{pressure_diff}
\ee
Due to the linearity of \eqs{incompressible}{stokes}, one expects a linear relation
\be
\left(\begin{array}{c}
f_1\\f_2
\end{array}\right)
= 4\mu q\, {\bf L}
\left(\begin{array}{c}
v_1\\v_2
\end{array}\right)
\ee
between normal forces applied on membranes $f_{1,2}$ and the $z$-component of flow velocity at the membranes, $v_{1,2}$, which defines the resistance matrix $4\mu q{\bf L}$. 
Following Brochard and Lennon~\cite{brochard} and Seifert~\cite{seifert_dynamics}, one seeks for solutions of the type $v_\parallel = v_x(x,z) = f(z)e^{iqx - \gamma t}$, and $v_z(x,z)=g(z)e^{iqx - \gamma t}$. From \eqs{incompressible}{noslip} we find, by writing $\tilde z=qz$ and $\tilde\eta_1=q\eta_1$
\bea
g_1(\tilde z) &=& A_1\left(\sinh \tilde z-\tilde z\cosh \tilde z + \frac{\tilde\eta_1\sinh \tilde\eta_1 \tilde z\sinh \tilde z}{\sinh \tilde\eta_1+\tilde\eta_1\cosh \tilde\eta_1}\right)\\
p_1 (\tilde z) &=& 2\mu q A_1\left(\frac{\tilde\eta_1\sinh \tilde\eta_1 \sinh \tilde z}{\sinh \tilde\eta_1+\tilde\eta_1\cosh \tilde\eta_1}-\cosh \tilde z\right)\\
g_2(\tilde z) &=& (A_2+B_2\tilde z)e^{\tilde z}+(C_2+D_2\tilde z)e^{-\tilde z}\\
p_2(\tilde z) &=& 2\mu q (B_2e^{\tilde z}+D_2e^{-\tilde z})\\
g_3(\tilde z) &=& (C_3+D_3\tilde z)e^{-\tilde z}\\
p_3(\tilde z) &=& 2\mu q D_3e^{-\tilde z}
\eea
Finally by solving the system of four equations \eqs{continuity}{incomp_memb}, we determine the four coefficients $A_2,B_2,C_2$ and $D_2$ and inserting the result in \eq{pressure_diff}, we find
\be
{\bf L}(\tilde\eta_1,\tilde\delta) = {\bf A}(\tilde\eta_1) +{\bf B}(\tilde\delta) 
\label{resistance_matrix}
\ee
where $\tilde \delta=q\delta=q(\eta_2-\eta_1)$ ($\delta=\eta_2-\eta_1$ is the average distance between membranes),
\be
{\bf A}(\tilde\eta_1) =  \frac12 \left(\begin{array}{cc}
\frac{\tilde\eta_1+\cosh \tilde\eta_1\sinh \tilde\eta_1}{\sinh^2 \tilde\eta_1-\tilde\eta_1^2} -1 & 0 \\
0 & 0
\end{array}\right)
\ee
and
\be
{\bf B}(\tilde\delta) =  \frac12 \left(\begin{array}{cc}
1 + \frac{\tilde\delta +\cosh \tilde\delta\sinh \tilde\delta}{\sinh^2 \tilde\delta-\tilde\delta^2}  &  -\frac{\sinh \tilde\delta+\tilde\delta\cosh \tilde\delta}{\sinh^2 \tilde\delta-\tilde\delta^2} \\
 -\frac{\sinh \tilde\delta+\tilde\delta\cosh \tilde\delta}{\sinh^2 \tilde\delta-\tilde\delta^2} & 1 + \frac{\tilde\delta +\cosh \tilde\delta\sinh \tilde\delta}{\sinh^2 \tilde\delta-\tilde\delta^2} 
\end{array}\right).
\ee
The damping matrix in \eq{membrane_dynamics} is nothing but the inverse of the resistance matrix
\be
 \mbox{\boldmath${\Lambda}$}(q,\eta_1,\eta_2) = \frac1{4\mu q} {\bf L}^{-1}(q\eta_1,q\delta).
\ee
The resistance matrix $4\mu q{\bf A}(\tilde\eta_1)$ corresponds to the fluid friction induced by the substrate where the no-slip boundary condition applies, whereas the symmetric resistance matrix $4\mu q{\bf B}(\tilde \delta)$ corresponds to the mutual hydrodynamic interactions between both membranes.

In the limit $\tilde \eta_1\to\pm\infty$, the fluid friction disappears (${\bf A}\to0$), and we are left with the symmetric resistance matrix $4\mu q{\bf B}(\tilde \delta)$, the large and low $q$ limits of which were studied by Brochard and Lennon~\cite{brochard,nelson}. On the other hand, in the limit $\tilde\delta\to\infty$, membranes decouple since ${\bf B}\to{\bf I}_2$ and we find the result of Seifert for a single membrane close to a substrate~\cite{seifert_dynamics}. This is also the case in the limits $\tilde\eta_1\to 0$ or $\tilde\delta\to0$ where the matrix ${\bf L}$ becomes a scalar
\be
L(x)= \frac12 \left(1+\frac{x+\cosh x\sinh x}{\sinh^2 x-x^2}\right)
\label{L_single}
\ee
where $x=\tilde\delta$ or $x=\tilde\eta_1$ respectively. The linearized lubrication approximation $x\ll1$ in \eq{L_single}, i.e. $L(x)=3/x^3$, corresponds to a flow parallel to the substrate~\cite{prost}.

\subsection{Damping rates}

Solving the coupled Langevin equations \eq{membrane_dynamics} amounts to finding the eigenvalues and the  eigenmodes of the damping matrix  $\mbox{\boldmath${\gamma}$}(q) = \mbox{\boldmath${\Lambda}$}(q) {\bf \hat U}(q)$ where 
\be
{\bf \hat U}(q)\equiv \left(\begin{array}{cc}
\sigma q^2 + \kappa q^4+ A_1 + A_2 &  - A_2 \\
-A_2  &  \sigma q^2 + \kappa q^4+A_2 
\end{array}\right).
\ee
By writing the two eigenvalues $\gamma_1(q)$ and $\gamma_2(q)$,  \eq{membrane_dynamics} becomes, in the eigenbasis
\be
\frac{\partial w_i(q,t)}{\partial t} = -\gamma_i (q) w_i(q,t) + ({\bf R}^{-1})_{ij}\zeta_j(q,t)
\label{membrane_dynamics2}
\ee
where ${\bf R}$ is the transformation matrix, ${\bf u}={\bf R} {\bf w}$. 
The solution for the temporal correlation function of $w_i(q,t)$ is
\be
\langle w_i(q,t)w_i(-q,0)\rangle =  \langle w_i(q,0)^2\rangle e^{-\gamma_i(q) t}
\ee
and the correlation function $\langle u_i(q,t) u_i(-q,0) \rangle$ can then be computed coming back to the initial coordinates and using
\bea
&&\langle|u_2(q)|^2\rangle = \frac{\sin^2\phi}{\lambda_1+\sigma q^2 +\kappa q^4} + \frac{\cos^2\phi}{\lambda_2+\sigma q^2 +\kappa q^4}\\
&&\langle u_2(q)u_1(-q)\rangle = \sin\phi\cos\phi \left(\frac1{\lambda_2+\sigma q^2 +\kappa q^4} \right.\nonumber\\ &&\left.- \frac1{\lambda_1+\sigma q^2 +\kappa q^4}\right).
\eea

\begin{figure}[t]
\begin{center}
\includegraphics[width=8cm]{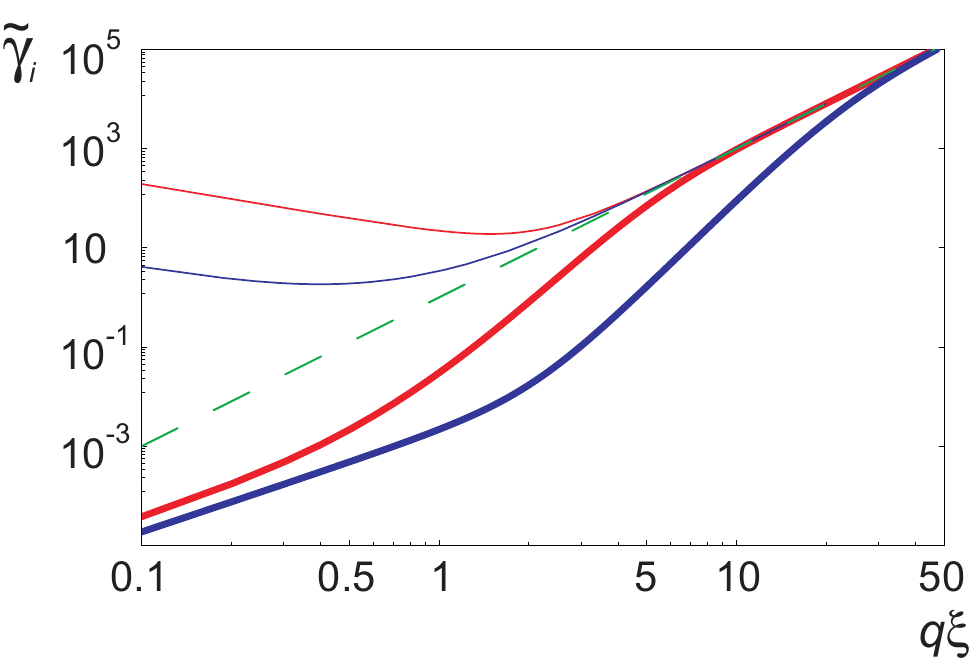}
\caption{Log-log plot of the (dimensionless) damping rates $\tilde\gamma_i$ versus adimensional wave-vector $q\xi$ for $\Xi_1=30$ and $\Xi_2=3$ at temperature $T=1.5\,\theta$ ($d_1=2$~nm and $d_2=3$~nm). Thick solid lines correspond to the full damping rates $\tilde \gamma_i$ ($i=1$ in red and 2 in blue) whereas thin solid lines are damping rates without hydrodynamics \eq{damping_noHI}, and the dashed line is the asymptotic result $(q\xi)^3$.}
\label{damping_rates}
\end{center}
\end{figure}

The eigenvalues $\gamma_i(q)$ are increasing functions (see \ref{damping_rates}) where limiting forms  for $q\to 0$ show a quadratic behaviour
\be
\gamma_1(q) = \frac{A_2 \delta^3}{12\mu}q^2 \quad \mathrm{and}\quad
\gamma_2(q) = \frac{A_1\eta_1^3}{12\mu}q^2.
\label{damping_0}
\ee
For $q\to\infty$ membranes decouple and we simply recover the free membrane damping rate \eq{damping_single}
\be
\gamma_1(q) = \gamma_2(q) = \frac{\kappa}{4\mu}q^3.
\ee
Note that without taking into account hydrodynamics, the damping rates $\gamma_i(q)$ are
\be
\gamma_i^{\rm woHI} (q)=\frac{\kappa q^4+\sigma q^2+\lambda_i}{4\mu q}
\label{damping_noHI}
\ee
where $\lambda_i$ are the eigenvalues given in \eq{lambda12}. Hence hydrodynamics induce a large decrease of the damping rates for small and intermediate values of $q$, which is expected due to the no-slip condition for the flow velocity at the substrate that slows down hydrodynamics.

Introducing the elastic decay length
\be
\xi\equiv\sqrt{\frac{2\kappa}{\sigma}}
\ee 
allows us to define the relevant dimensionless parameters in the dimensionless damping rates $\tilde \gamma_i$ defined as
\be
\tilde \gamma_i\left(q\xi, \tilde A_1, \tilde A_2, \frac{\eta_1}{\xi}, \frac{\eta_2}{\xi}\right)=\frac{4\mu \xi^3}{\kappa}\gamma_i(q)
\label{adim_damping}
\ee
where the $\tilde A_i$ and $\eta_i$ are fixed for a given couple of $(\Xi_1,\Xi_2)$.

Damping rates \eq{adim_damping} are plotted in \ref{damping_rates} for $(\Xi_1,\Xi_2)=(30,3)$ at relative temperature $T/\theta=1.5$. Following the previous Section, it corresponds to variational parameters equal to $\tilde A_1=18.6$ and $\tilde A_2= 0.4$. From the phase diagram in \ref{2memb}, we are thus in region~(1) close to the unbinding of membrane~2 [$\alpha(\eta_1-d_1)=0.72$ and $\alpha(\eta_2-d_2)=4.76$]. However by choosing $d_1=2$~nm and $d_2=3$~nm, this yields $\eta_1/\xi=0.07$ and $\delta/\xi=0.3$, and the substrate--membrane~1 and inter-membrane distances are small enough to induce a sensible slowing down at small and intermediate wave-vectors $q$. Indeed, when hydrodynamic interactions are neglected (thin lines in \ref{damping_rates}) damping rates are larger  by 4 to 6 orders of magnitude: for $q\to0$, $\gamma_i^{\rm woHI} (q)$ reduces to $\lambda_i/4\mu q$ which should be compared to actual values in \eq{damping_0}. Moreover, the damping rates of the two normal modes are less different for low $q$ than expected without hydrodynamics. From \eq{damping_0} the low $q$ values of the damping rates $\gamma_i(q)$ are controlled by $A_1\eta_1^3$ for mode~1 and $A_2\delta^3$ for mode~2. Usually the values of these quantities are quite close, which can be explained by the fact that an increase of $A_i$ implies a decrease of $\eta_i$ as shown in the previous Section [see for instance \eqs{Ai}{ui}],  and the decoupling of the two modes occurs only at intermediate $q$ (see \ref{damping_rates}).

At room temperature, the values used in \ref{damping_rates} yields $\xi=56$~nm (for $\sigma=7\times10^{-5}$~J.m$^{-2}$ and $\kappa=30 k_BT$) with $\alpha_1^{-1}=\alpha_2^{-1}=2.5$~nm, $D_1\simeq10^{-6}$~J.m$^{-2}$, and $D_2\simeq10^{-7}$~J.m$^{-2}$ which are reasonable values~\cite{daillant,lin}. 
Note that close to the unbinding transition for $T\lessapprox T_u$, the average square value of height functions [\eq{correlations3}] takes large values, the approximation of almost planar membrane fails, and dynamical renormalization techniques should be applied.


\section{Discussion}


The variational approach that we have developed to describe the statistical physics of double supported membranes is a complementary approach to numerical Monte-Carlo simulations and numerical transfer matrix methods. It allows an analytical determination of the unbinding temperatures, and the variational parameters, the effective spring constants $A_i(T)$ and the substrate-membrane and inter-membrane distances $\eta_i(T)$, at any temperature and any values of the dimensionless adsorption strengths $(\Xi_1,\Xi_2)$. 

This is central for analyzing X-ray specular reflectivity spectra since the $A_i(T)$ are necessary to fit the experimental structure factors. In the experiments by Daillant \textit{et al.}~\cite {daillant}, the measured effective spring constant of the upper bilayer $A_2$, which has a large effect on the spectrum shape, was found surprisingly weak, $A_2\approx 5\times10^{10}$~J.m$^{-4}$,  much lower than the bare spring constant of the potential estimated at $2\alpha^2 D_2\approx2\times10^{14}$~J.m$^{-4}$. With these values we find $\tilde A_2/\Xi_2\simeq 2\times10^{-4}$, a value extremely low suggesting that the system is closed to the unbinding transition for the upper membrane. Inserting their fitted parameter values for the surface tension and the bending rigidity ($\sigma=5\times10^{-4}$~J.m$^{-2}$ and $\kappa\simeq 1.5\times10^{-19}$~J), we find $\Xi_2\simeq200$. Assuming for the first membrane $\Xi_1\simeq100$, the phase diagram \ref{2memb}(b) shows that this would lead to $T\simeq 5\theta$. With $\alpha\simeq 2$~nm$^{-1}$ we thus find $T\simeq 300$~K which is roughly the temperature of the experiment. Although this is a crude estimation, we find the correct values, even if the  parameters $A_i$ depend in a complex manner on both inter-membrane and substrate-membrane potentials. Note that more generally, even if the transition is not perfectly described within our variational approach, the error made on the value of the unbinding temperature $T_u$ is small since $A_i(T)$ varies rapidly close to $T_u$.

In recent experiments by Stidder \textit{et al.}~\cite{stidder}, unbinding of the upper membrane in DPPE double stacked bilayers with 10 mol\% of cholesterol has been observed between 52.2~$^\circ$C and 56.4~$^\circ$C by neutron reflectivity. Their results suggest i) a discontinuous unbinding transition since they observe both an hysteresis and a larger but \textit{finite} roughness, $\sqrt{\langle u^2\rangle}= 0.6\pm1.2$~nm, at the transition, and ii) that the presence of 10 mol\% of cholesterol modifies the direct interaction potential and decreases the unbinding temperature. We qualitatively obtain such unbinding transitions for $\Xi_2\lessapprox 2\Xi_1$ but values of the surface tension and the direct adsorption potential are needed, to do a quantitative comparison. Nevertheless, these experiments allow one to hope that a detailed characterization of double lipid bilayers will be achieved experimentally in the very near future.

When compared to numerical results of Refs.~\cite{netz,netz2} where only the case $\Xi_1=\Xi_2$ was explored, we also find sequential thermal unbinding transitions: for $\Xi_1=\Xi_2=30$ the upper bilayer unbinds at $T_u(2)/\theta= 2.6$ and the remaining bilayer then unbinds at $T_u(1)/\theta=3.3$ (see \ref{figlambda} and~\ref{1membplus}). Moreover,  membrane fluctuations increase when moving away from the substrate, $\eta_2-d_2>\eta_1-d_1$. We show that this behaviour is valid whenever $\Xi_2<2\Xi_1$. However, for higher values of $\Xi_2$, one observe the reverse case where the stack of two membranes unbinds as a whole. This has been shown for bundles of strings in two dimensions~\cite{hiergeist}. Our phase diagram (\ref{2memb}) resembles the one of Ref.~\cite{hiergeist} for strings. A way to compare them is to change the coordinates of the phase diagram into $\Xi_{1,2}(T/\theta)^2\propto T^2/(\kappa D_{1,2})$ in order to eliminate the surface tension (which does not exist in the string problem). Note that, for strings, the parameter $\kappa$ becomes an elastic (stretching) parameter. The qualitative diagram is then similar with a region corresponding to the peeling process and another one where the bundle desorbs as a whole. However, contrary to these studies, we find a range of parameters $\Xi_2$ for which the double supported membranes de-stack completely at the transition, this region being larger and larger when $\Xi_2$ increases. The richness of the phase diagram is related to the fact that, contrary to Refs.~\cite{netz,netz2,hiergeist} we consider bilayers with a microscopic surface tension, which is the more general case.

In this work, we assumed the substrate to be flat and did not considered the substrate roughness which leads to pinned or hovering binding states for the first membrane~\cite{hansen}. How does the influence of the quenched substrate disorder propagate to the second membrane is an important issue. One might suggest that this effect for the upper membrane is somehow annealed by thermal fluctuations of the lower one, and the pinning effect is then less pronounced. Anyway, the influence of substrate roughness is central from an experimental point of view and deserves a quantitative study.

From an experimental perspective, we now discuss the biophysical
interest of designing supported double (or even multiple) lipid
bilayers. Ideally, for the upper bilayer to deserve to be
qualified as ``nearly floating'' and weakly perturbed by the substrate,
both $\delta$ and $1/A_2$ must be large so that this bilayer
fluctuates nearly freely, while still being conveniently observable by
spectroscopy techniques. One can in principle obtain a single
adsorbed bilayer with the same properties, by approaching its
unbinding transition. However, when the
experimental goal is to model a biological situation (e.g. plasma
membrane or cell-cell junction), both membrane composition and
temperature are imposed by the biological context. This reduces the
possibility to design a single ``nearly floating'' bilayer. By contrast,
our study suggests that such a regime can be reached by stacking two (or more)
bilayers, because thermal fluctuations are enhanced in the upper
one. Our goal is to help to anticipate in which regime of parameters 
this situation is likely to occur. 

Another experimental interest of studying thermal fluctuations, both
at the equilibrium and dynamical levels, lies in the possibility to
study molecular diffusion in supported bilayers, for example by single 
molecule tracking. On the one hand, fluctuations affect apparent diffusion 
coefficients because the projected area is different from the real membrane one; 
the relationship between apparent and real diffusion coefficients depends 
on the relative timescales of membrane fluctuations and diffusion~\cite{naji1}. 
On the other hand, it has been recently shown by numerical simulations~\cite{naji2} that membrane elasticity decrease protein mobility, essentially due to the viscous drag of the deformed membrane around the protein, but that membrane fluctuations slightly increase protein mobility~\cite{reister}. However such effect is weak compared to viscous losses~\cite{naji2}. 
It will be useful to quantify these effects, related to the amplitude of fluctuations, in the case of stacked bilayers. This work is currently in progress~\cite{progress}.

\acknowledgements
We thank Laurence Salom\'e for her precious knowledge of the experimental context and John Palmeri for illuminating discussions.

\appendix
\section{Correlation functions for double supported bilayer}
\label{correlation_functions}

By performing an inverse Fourier transform of \eq{correlations} and coming back to initial coordinates, one finds the following height-height correlation functions
\bea
C_{11}(\br) &=& \langle u_1(\br) u_1({\bf 0})\rangle\nonumber\\ &=& \frac{k_BT}{2\pi}\left[\cos^2\phi \int_0^\infty \mathrm{d}q \frac{q J_0(q|\br|)}{\lambda_1+\sigma q^2+\kappa q^4}\right.\nonumber\\&& \left.+\sin^2\phi \int_0^\infty \mathrm{d}q \frac{q J_0(q|\br|)}{\lambda_2+\sigma q^2+\kappa q^4}\right]\\
C_{12}(\br) &=& \langle u_1(\br) u_2({\bf 0})\rangle\nonumber\\ &=&  \frac{k_BT\cos\phi\sin\phi}{2\pi}\left[ -\int_0^\infty \mathrm{d}q \frac{q J_0(q|\br|)}{\lambda_1+\sigma q^2+\kappa q^4}\right.\nonumber\\&&\left.+\int_0^\infty \mathrm{d}q \frac{q J_0(q|\br|)}{\lambda_2+\sigma q^2+\kappa q^4}\right]\\
C_{22}(\br) &=& \langle u_2(\br) u_2({\bf 0})\rangle\nonumber\\ &=& \frac{k_BT}{2\pi}\left[\sin^2\phi \int_0^\infty \mathrm{d}q \frac{q J_0(q|\br|)}{\lambda_1+\sigma q^2+\kappa q^4}\right.\nonumber\\&&\left.+\cos^2\phi \int_0^\infty \mathrm{d}q \frac{q J_0(q|\br|)}{\lambda_2+\sigma q^2+\kappa q^4}\right]
\label{correlations2}
\eea
where $J_0$ is the Bessel function of the first kind, and both $\phi$ and $\lambda_i$ depend on the variational parameters $A_1(T)$ and $A_2(T)$, following Eqs.~(\ref{lambda12}) and (\ref{theta}).

\section{Stack of $n$ supported membranes}
\label{n_membranes}

In the case of $n$ stacked membranes on a substrate at $z=0$,
the full Hamiltonian can be written as the tensor product of the usual Helfrich Hamiltonian of a single fluctuating bilayer membrane and the Hamiltonian which describes interactions between membranes:
\be
\mathcal{H}[\{h_j(\br)\}] = \mathcal{H}_{\rm Helfrich}[h(\br)]\otimes\mathbf{1}_{\rm phonon}+\mathbf{1}_{\rm Helfrich}\otimes\mathcal{H}_{\rm phonon}[h_j]
\label{H}
\ee
where $ \mathcal{H}_{\rm Helfrich}$ is the classical Helfrich Hamiltonian of a free membrane, \eq{helfrichH} with $V=0$.
The phonon Hamiltonian is described by a succession of harmonic springs in the $z$-direction
\be
\mathcal{H}_{\rm phonon}[h_j]= \frac12 \sum_{j=0}^{n-1} A_j \left[h_{j+1}(\br)-h_j(\br)\right]^2.
\label{phononH}
\ee
Taking thermal expansion into account would require a variational approach where the $A_i$ are different and computed variationally from, for instance, Morse potentials as defined in \eq{Morse}. However, even if possible in principle, this approach is hardly tractable in practice. In the following of this appendix we consider all the $A_i$ equal, $A_i=A$, $\forall\, i$.

Since the boundary conditions for the phonon modes are: $h_0(\br)=0$ and the last membrane free, we find from Eq.~(\ref{phononH}) the following eigenmodes and eigenvalues
\bea
h_j^\ell &=&-\frac2{\sqrt{2n+1}}\sin\left(\frac{2\ell+1}{2n+1}\pi\,j\right)\\
\lambda_\ell &=& 4\sin^2\left(\frac{2\ell+1}{2n+1}\frac\pi2\right)
\eea
with $0\leq\ell\leq n-1$. For $n=2$ we find $\lambda_0=4\sin^2(\pi/10)$ and $\lambda_1=4\sin^2(3\pi/10)$ which corresponds to $\lambda_2/A_2$ and $\lambda_1/A_1$ respectively, in \eq{lambda12} (with $A_1=A_2=A$).

The Hamiltonian~(\ref{H}) is diagonal in the Fourier $(\ell,\bq)$ space, where $\ell$ is the index for the deformation modes in the $z$-direction and $\bq$ is the planar wave-vector in the ($x,y$) plane:
\be
\mathcal{H}[\hat{h}(\ell,\bq)]=\frac12 \int \frac{\mathrm{d}^2\bq}{(2\pi)^2} \sum_{\ell=0}^{n-1}(A\lambda_\ell+\sigma\bq^2+\kappa\bq^4)|\hat{h}(\ell,\bq)|^2
\ee
where we define the Fourier transform of height fluctuations $h_j(\br)$, by
\be
\hat{h}(\ell,\bq)=\frac2{\sqrt{2n+1}}\sum_{j=0}^n \sin\left(\frac{2\ell+1}{2n+1}\pi\,j\right)  \int \mathrm{d}^2\br \, e^{-i\br\cdot\bq} \,h_j(\br)
\ee
The Hamiltonian being Gaussian, the height-height correlation function written in Fourier space is
\be
\langle\hat{h}(\ell,\bq)\hat{h}(\ell',\bq')\rangle=(2\pi)^2\delta_{\ell,\ell'}\delta(\bq+\bq')\frac{k_BT}{A\lambda_\ell+\sigma\bq^2+\kappa\bq^4}.
\ee
The height-height correlation function $C_{jk}(\br)=\langle h_j(\br) h_k({\bf 0})\rangle$ between membrane $j$ at $\br$ and membrane $k$ at $\br={\bf 0}$ is defined as
\bea
C_{jk}(\br) &=& \frac2\pi \frac{k_BT}{2n+1}\sum_{\ell=0}^{n-1}  \sin\left(\frac{2\ell+1}{2n+1}\pi j\right) \sin\left(\frac{2\ell+1}{2n+1}\pi k\right) \nonumber\\&&\times \int \mathrm{d}q \frac{q J_0(q|\br|)}{A\lambda_\ell+\sigma q^2+\kappa q^4}.
\eea
This last equation is thus  the generalization of \eq{correlations2} for $n$ membranes.

Similar but different results have been obtained in the contexts of smectic-A films~\cite{romanov} with a different boundary condition for the layer $n$, and of solid-supported multilayers~\cite{constantin} using a continuous description of Eq.~(\ref{phononH}) valid for large $n$ (in the continuum limit $d\to 0$, $n\to \infty$ with $L=nd$ held fixed).

\end{document}